\documentclass[aps,twocolumn,prb]{revtex4-1}%
\usepackage{amsfonts,amsmath,amssymb,graphicx,natbib}
\usepackage{array}
\usepackage{bm}
\usepackage{mathrsfs}
\usepackage{hyperref} 
\hypersetup{ 
 backref=true,    
 pagebackref=true,
 hyperindex=true, 
 colorlinks=true, 
 breaklinks=true, 
 urlcolor= blue,  
 linkcolor= [rgb]{0.7,0,0},
 citecolor= [rgb]{0,0.5,0},
 bookmarks=true,  
 bookmarksopen=true,            
 pdftitle={quantum cluster methods}, 
 pdfauthor={David SENECHAL},     
 pdfsubject={Physics}          
}

\newcommand{\InsertFigure}[3]{%
\begin{figure}
\begin{center}
\includegraphics[width=#3]{figs/#2.pdf}
\caption{#1}
\end{center}
\end{figure}
}

\newcolumntype{C}{>{$}c<{$}}
\newcolumntype{R}{>{$}r<{$}}
\newcolumntype{L}{>{$}l<{$}}

\newcommand{\BEQ}{\begin{equation}}
\newcommand{\EEQ}{\end{equation}}
\newcommand{\ALIGN}[1]{\begin{align}#1\end{align}}
\newcommand{\SPLIT}[1]{\begin{equation}\begin{split}#1\end{split}\end{equation}}
\newcommand{\ALIGNED}[1]{\begin{aligned}#1\end{aligned}}
\newcommand{\BEA}{\begin{eqnarray}}
\newcommand{\EEA}{\end{eqnarray}}
\newcommand{\BA}{\begin{array}}
\newcommand{\EA}{\end{array}}

\newcommand{\R}{\rangle}
\renewcommand{\L}{\langle}
\renewcommand{\r}{\right}
\renewcommand{\l}{\left}

\newcommand{\hf}{\frac12}

\newcommand{\Frac}[2]{\frac{\displaystyle #1}{\displaystyle #2}}

\newcommand{\pdel}[2]{\Frac{\partial #1}{\partial #2}}

\newcommand{\Quad}{\qquad\qquad}
\newcommand{\up}{\uparrow}
\newcommand{\dn}{\downarrow}
\newcommand{\dg}{\dagger}
\renewcommand{\d}{\partial}
\newcommand{\id}{\mathbf{1}}

\newcommand{\G}{\Gamma}
\newcommand{\MATRIX}[1]{\begin{pmatrix}#1\end{pmatrix}}
\newcommand{\varia}[2]{\frac{\delta #1}{\delta #2}}
\newcommand{\Hc}{\mathscr{H}}
\newcommand{\Gc}{\mathscr{G}}
\newcommand{\Gcv}{{\Gc\kern-0.76em\Gc}}
\newcommand{\Oc}{\mathcal{O}}
\newcommand{\Nc}{\mathscr{N}}
\newcommand{\Gg}{\mathfrak{G}}

\newcommand{\cpp}{C{$\scriptstyle ++$}}

\newcommand{\er}{\mathrm{e}}
\newcommand{\dr}{\mathrm{d}}
\newcommand{\tr}{\,\mathrm{tr}}
\newcommand{\Tr}{\,\mathrm{Tr}}
\newcommand{\Text}[1]{~\text{#1}~}
\renewcommand{\Im}{\,\mathrm{Im}}
\renewcommand{\Re}{\,\mathrm{Re}}

\newcommand{\Dcv}{\mathscr{D}\kern-1.9ex\mathscr{D}}
\newcommand{\Ec}{\mathscr{E}}

\DeclareMathAlphabet{\mathitbf}{OML}{zplm}{b}{it}
\newcommand{\vectori}[1]{\mathitbf{#1}}

\renewcommand{\vector}[1]{\vectori{#1}}

\newcommand{\Dv}{\vector{D}}
\newcommand{\epsv}{\vectori{\varepsilon}}
\newcommand{\ev}{\vector{e}}

\newcommand{\Gammav}{\vector{\Gamma}}
\newcommand{\gv}{\vector{g}}
\newcommand{\Gv}{\vector{G}}

\newcommand{\kv}{\vector{k}}
\newcommand{\Kv}{\vector{K}}
\newcommand{\Lambdav}{\vectori{\Lambda}}

\newcommand{\Lv}{\vector{L}}

\newcommand{\Mv}{\vector{M}}

\newcommand{\Pv}{\vector{P}}

\newcommand{\Qv}{\vector{Q}}
\newcommand{\rv}{\vector{r}}
\newcommand{\Rv}{\vector{R}}
\newcommand{\sv}{\vector{s}}
\newcommand{\Sigmav}{\vectori{\Sigma}}
\newcommand{\Sv}{\vector{S}}
\newcommand{\thetav}{\vectori{\theta}}
\newcommand{\tv}{\vector{t}}
\newcommand{\Tv}{\vector{T}}
\newcommand{\Uv}{\vector{U}}

\newcommand{\Vv}{\vector{V}}

\newcommand{\xiv}{\bm{\xi}}

\newcommand{\pvec}{\wedge}

\renewcommand{\a}{\alpha}
\renewcommand{\b}{\beta}
\newcommand{\g}{\gamma}

\newcommand{\eps}{\varepsilon}
\newcommand{\om}{\omega}
\newcommand{\Om}{\Omega}
\newcommand{\s}{\sigma}
\renewcommand{\t}{\theta}

\newcommand{\kt}{{\tilde k}}
\newcommand{\kvt}{{\tilde\kv}}
\newcommand{\rvt}{{\tilde\rv}}

\newcommand{\BI}{\begin{itemize}\renewcommand{\labelitemi}{$\bullet$}\renewcommand{\labelitemii}{--}}
\newcommand{\EI}{\end{itemize}}
\newcommand{\BE}{\begin{enumerate}}
\newcommand{\EE}{\end{enumerate}}

\begin{document}
\title{An introduction to quantum cluster methods\\[8pt]
\rm Lectures given at the CIFAR - PITP International Summer School on Numerical Methods for Correlated Systems in Condensed Matter, Sherbrooke, Canada, May 26 -- June 6 2008}
\author{David S\'{e}n\'{e}chal}
\affiliation{D\'{e}partement de physique and Regroupement qu\'{e}b\'{e}cois sur les mat\'{e}riaux de pointe, Universit\'{e} de Sherbrooke, 
Sherbrooke, Qu\'{e}bec, Canada, J1K 2R1}
\date{May 2008}

\begin{abstract}
These lecture notes provide an introduction to quantum cluster methods for strongly correlated systems.
Cluster Perturbation Theory (CPT), the Variational Cluster Approximation (VCA) and Cellular Dynamical Mean Field Theory (CDMFT) are described, as well as the exact diagonalization solver for the cluster. 
Potthoff's self-energy functional formalism is reviewed.
Some numerical procedures, in particular regarding the exact diagonalization method and the frequency-momentum integrals needed in VCA, are discussed in detail.
\end{abstract}

\maketitle
\section{Introduction}

Classic numerical approaches to lattice models such as the Hubbard model are usually based on a solution of the model on a periodic lattice with a small number of sites.
For instance, Exact Diagonalizations (ED) are performed on periodic systems with no more than $\sim 20$ sites and Quantum Monte Carlo (QMC) is limited in practice on systems with $\lesssim 100$ sites.
Some extrapolation is then needed to make statements about the thermodynamic (e.g. infinite-size) limit.
One advantage of such approaches is their relative simplicity and lack of ambiguity.
A disadvantage is that broken symmetry states need careful analysis to be identified, as they are fully revealed only in the thermodynamic limit.

Quantum cluster methods are a set of closely related approaches that consider instead a finite cluster of sites embedded in the infinite lattice.
The embedding is done by adding to the cluster additional fields or bath degrees of freedom such as to best represent the effect of the surrounding infinite lattice.
Variational or self-consistency principles are used to set the values of these additional parameters.
In these approaches, broken symmetry states can appear even for the smallest clusters used, somewhat like ordinary mean-field theory.
However, unlike mean field theory, these approaches are dynamical and retain the full effect of strong correlations.
These methods are usually known by their acronyms:
\BE
\item VCA (Variational Cluster Approximation)\cite{Potthoff:2003} or VCPT (Variational Cluster Perturbation Theory)
\item CDMFT (Cluster/Cellular Dynamical Mean-Field Theory)\cite{Kotliar:2001}
\item DCA (Dynamical Cluster Approximation)\cite{Hettler:1998,Hettler:2000}
\EE
The first two of these methods (VCA and CDMFT) can be understood within a more general framework called the Self-Energy Functional Approach (SFA), proposed by M.~Potthoff.\cite{Potthoff:2003b}
The last one (DCA) cannot, as usually formulated, but is a momentum-space analog of CDMFT.
Both DCA and CDMFT are cluster generalizations of Dynamical Mean Field Theory (DMFT).\cite{Georges:1992, Jarrell:1992}

These lecture notes will concentrate on VCA, its precursor CPT, and CDMFT, along with the exact diagonalization solver.
DCA will be discussed by M.~Jarrel, later during this school.
Readers are referred to the excellent review by Maier {\it et al.}\cite{Maier:2005} for different aspects of cluster methods, including alternate solvers (that review, however, was written before the VCA technique was mature).

Each of these cluster methods is in turn dependent on a solution of the cluster Hamiltonian $H'$ -- which differs from the lattice Hamiltonian $H$ -- by a number of different (exact or approximate) methods.
The cluster being often compared to an impurity, we often refer to these as different \textit{impurity solvers}, although the expression \textit{cluster solver} is more appropriate.
In these notes, we will describe in some detail a solver based on exact diagonalizations.

We will be concerned with the one-band Hubbard model, defined on a lattice $\g$ whose sites will be labelled by position vectors $(\rv,\rv',\dots)$.
The destruction operator for an electron on the Wannier orbital centered at the site $\rv$ with spin $\s$ will be denoted $c_{\rv\s}$, and the corresponding number operator will be $n_{\rv\s}$.
With this notation, the lattice Hamiltonian reads
\BEQ\label{eq:Hubbard}
H = \sum_{\rv,\rv', \s} t_{\rv\rv'} c^\dg_{\rv\s}c_{\rv'\s} + U\sum_{i}n_{\rv\up}n_{\rv\dn} - \mu\sum_\rv n_\rv
\EEQ
where $t_{\rv\rv'}$ is the hopping matrix, $U$ is the one-site Coulomb repulsion and $\mu$ is the chemical potential, which we find convenient to include in the Hamiltonian.
We will assume, for counting purposes, that the lattice $\g$ is periodic, with a large (i.e., billions) but finite number of sites $N$.
multi-band Hubbard models are a simple extension of this, and we can always keep in mind that the index $\s$ stands for both spin and band if we like.

This paper is organized as follows:
\BE
\item Section 2 reviews Cluster Perturbation Theory (CPT), the simplest of all cluster approaches, which is the basis of VCA and serves as a general introduction to cluster kinematics.
\item Section 3 reviews the Exact Diagonalization technique for computing the cluster's ground state and Green function, making use of the Lanczos and Band Lanczos methods.
\item Section 4 reviews Potthoff's self-energy functional approach, necessary to understand VCA (Section 5) and CDMFT (Section 6).
\item Roughly a third of this paper consists of appendices that explain some specific points in more detail.
In particular, Appendix A deals with cluster kinematics and is required reading before Section~2.
\EE

\section{Cluster perturbation theory}

\InsertFigure{A 10-site cluster and the corresponding superlattice vectors.\label{fig:B10}}{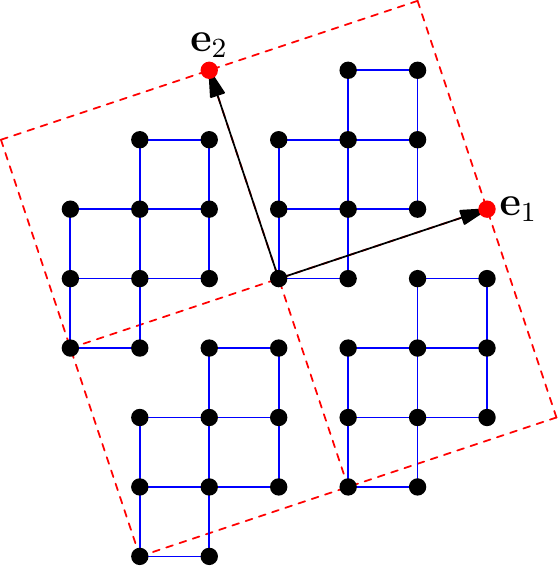}{0.65\hsize}

The simplest quantum cluster method is Cluster Perturbation Theory (CPT).\cite{Gros:1993, Senechal:2000}
CPT can be viewed as a cluster extension of strong-coupling perturbation theory\cite{Pairault:1998}, although limited to lowest order.\cite{Senechal:2002}
Its kinematical features are found in more sophisticated approaches like VCA or CDMFT.
The reader is strongly encouraged to read Appendix A, where much of the notation about clusters and indices is explained.

CPT proceeds as follows.
First a cluster tiling is chosen (see, e.g., Fig.~\ref{fig:B10}).
Then the lattice Hamiltonian $H$ is written as $H = H' + V$, where $H'$ is the cluster Hamiltonian, obtained by severing the hopping terms between different clusters, and $V$ contains precisely those terms.
$V$ is treated as a perturbation.
It can be shown, by the techniques of strong-coupling perturbation theory\cite{Senechal:2000, Senechal:2002}, that the lowest-order result for the lattice Green function is
\BEQ\label{eq:CPT0}
\Gv^{-1}(\om) = \Gv'{}^{-1}(\om) - \Vv~,
\EEQ
where $\Vv$ is the matrix of inter-cluster hopping terms and $\Gv'(\om)$ the exact Green function of the cluster.
This formula deserves a more thorough description: $\Gv$, $\Gv'$ and $\Vv$ are  matrices in the space $E$ one-electron states.
This space is the tensor product $\g\otimes B$ of the lattice $\g$ by the space $B$ of band and spin states.
For the remainder of this section we will ignore $B$, i.e., band and spin indices. 
In terms of compound cluster/cluster-site indices $(\rvt,\Rv)$, $\Gv'$ is diagonal in $\rvt$ and identical for all clusters, whereas $\Vv$ is essentially off-diagonal in $\rvt$.
Because of translation invariance on the superlattice, the above formula is simpler in terms of reduced wavevectors, following a partial Fourier transform $\rvt\to\kvt$:
\BEQ\label{eq:CPT1}
\Gv^{-1}(\kvt,\om) = \Gv'{}^{-1}(\om) - \Vv(\kvt)~.
\EEQ
The matrices appearing in the above formula are now of order $L$ (the number of sites in the cluster), i.e., they are matrices in cluster sites $\Rv$ only.
$\Gv'$ is independent of $\kvt$, whereas $\Vv$ is frequency independent.

The basic CPT relation (\ref{eq:CPT1}) may also be expressed in terms of the self-energy $\Sigmav$ of the cluster Hamiltonian as
\BEQ\label{eq:CPT2}
\Gv^{-1}(\kvt,\om) = \Gv_0{}^{-1}(\kvt,\om) - \Sigmav(\om)~,
\EEQ
where $\Gv_0(\kvt,\om)$ is the Green function associated with the non-interacting part of the lattice Hamiltonian.
This follows simply from the relations
\ALIGN{
\Gv'{}^{-1} &= \om - \tv' - \Sigmav \\
\Gv_0{}^{-1} &= \om - \tv' - \Vv ~,
}
where $\tv'$ is the restriction to the cluster of the hopping matrix (chemical potential included).
It is in the form (\ref{eq:CPT2}) that CPT was first proposed \cite{Gros:1993}.

A supplemental ingredient to CPT is the periodization prescription, that provides a fully $\kv$-dependent Green function out of the mixed representation $G_{\Rv\Rv'}(\kvt,\om)$.
Indeed, the cluster decomposition breaks the original lattice translation symmetry of the model.
The Green function (\ref{eq:CPT1}) is not fully translation invariant.
This means that it is not diagonal when expressed in terms of wavevectors: $\Gv\to G(\kv,\kv')$.
Due to the residual superlattice translation invariance, however, $\kv'$ and $\kv$ must map to the same wavevector of the superlattice Brillouin zone (or reduced Brillouin zone) and differ by an element of the reciprocal superlattice.
The periodization procedure proposed in Ref.~\onlinecite{Senechal:2000} applies to the Green function itself:
\BEQ\label{eq:CPT3}
G_{\rm per.}(\kv,\om) = \frac1{L}\sum_{\Rv,\Rv'} \er^{-i\kv\cdot(\Rv-\Rv')}G_{\Rv\Rv'}(\kvt,\om)~.
\EEQ
Moreover, since the reduced zone $\kvt$ is taken from is immaterial, on may replace $\kvt$ by $\kv$ in the above formula (i.e. replacing $\kvt$ by $\kvt+\Kv$ yields the same result).
This periodization formula may be heuristically justified as follows.
In the $(\Kv,\kvt)$ basis, the matrix $\Gv$ has the following form:
\BEQ
G_{\Kv\Kv'}(\kvt,\om) = \frac1{L}\sum_{\Rv,\Rv'} \er^{-i(\Kv\cdot\Rv-\Kv'\cdot\Rv')}G_{\Rv\Rv'}(\kvt,\om)~.
\EEQ
This form can be further converted to the full wavevector basis $(\kv=\Kv+\kvt)$ by use of the unitary matrix $\Lambdav$ of Eq~(\ref{eq:lambdaMatrix}):
\ALIGN{
&G(\kvt+\Kv,\kvt+\Kv') = \l(\Lambdav(\kvt)\Gv\Lambdav^\dg(\kvt)\r)_{\Kv\Kv'}  \notag\\
&= \frac1{L^2}\sum_{\Rv,\Rv',\Kv_1,\Kv_1'} \er^{-i(\kvt+\Kv-\Kv_1)\cdot\Rv} \er^{i(\kvt+\Kv'-\Kv'_1)\cdot\Rv'}G_{\Kv_1\Kv'_1} \notag\\
&= \frac1{L}\sum_{\Rv,\Rv'} \er^{-i(\kvt+\Kv)\cdot\Rv} \er^{i(\kvt+\Kv')\cdot\Rv'}G_{\Rv\Rv'}(\kvt,\om)~.
}
The periodization prescription (\ref{eq:CPT3}) amounts to picking the diagonal piece of the Green function ($\kv=\kv'$) and discarding the rest.
This makes sense in as much as the density of states $N(\om)$ is the trace of the imaginary part of the Green function:
\ALIGN{
N(\om) &= -\frac2N\Im\tr\Gv(\om) = -\frac2N\Im\sum_\rv G_{\rv\rv}(\om) \notag\\
&= -\frac2N\Im\sum_\kv G(\kv,\om)~,
}
and the spectral function $A(\kv,\om)$, as a partial trace, involves only the diagonal part.
Indeed, it is a simple matter to show from the anticommutation relations that the frequency integral of the Green function is the unit matrix:
\BEQ
-2\Im\int{\dr\om\over 2\pi}~\Gv(\om) = \id~.
\EEQ
This being representation independent, it follows that the frequency integral of the imaginary part of the off-diagonal components of the Green function vanishes.

\begin{figure}
\centering
\includegraphics[width=8cm]{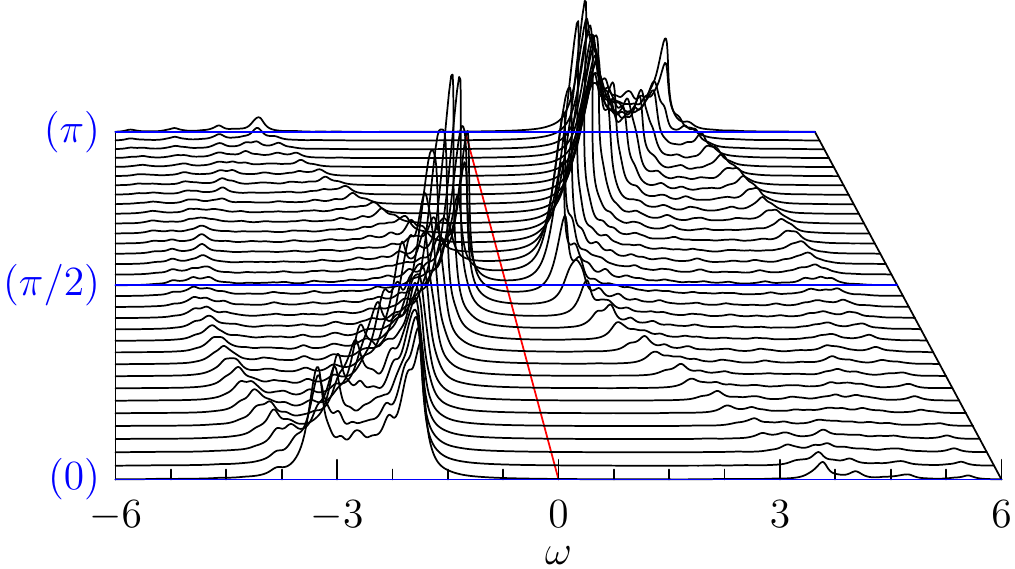}
\includegraphics[width=8cm]{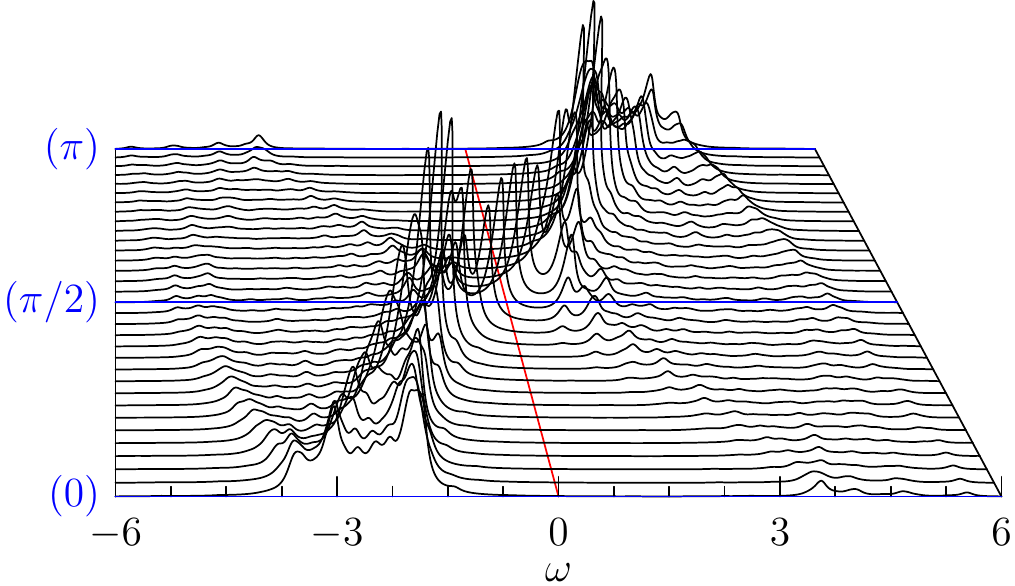}
\caption{Top: CPT spectral function of the one-dimensional, half-filled Hubbard model with $U=4$, $t=1$, with Green function periodization ($L=16$).
Bottom : the same, with Self-energy periodization instead; notice the important spectral weight
in the middle of the Mott gap.\label{fig:periodization}}
\end{figure}

Another possible prescription for periodization is to apply the above procedure to the self-energy $\Sigmav$ instead.
This is appealing since $\Sigmav$ is an irreducible quantity, as opposed to $\Gv$.
This amounts to throwing out the off-diagonal components of $\Sigmav$ before applying Dyson's equation to get $\Gv$, as opposed to discarding the off-diagonal part at the last step, once the matrix inversion towards $\Gv$ has taken place.
As Fig.~\ref{fig:periodization} shows, periodizing the Green function (Eq.~(\ref{eq:CPT3})) reproduces the expected feature of the spectral function of the one-dimensional Hubbard model.
In particular, the Mott gap that opens at arbitrary small $U$ (as known from the exact solution), whereas periodizing the self-energy leaves spectral weight within the Mott gap for abritrary large value of $U$.
Therefore we will always use Green function periodization.

CPT has the following characteristics:
\BE
\item Although it is derived using strong-coupling perturbation theory, it is exact in the $U\to0$ limit, as the self-energy disappear in that case.
\item It is also exact in the strong-coupling limit $t_{\rv\rv'}/U \to 0$.
\item It provides an approximate lattice Green function for arbitrary wavevectors.
Hence its usefulness in comparing with ARPES data.
Even though CPT does not have the self-consistency present in DMFT type approaches, at fixed computing resources it allows for the best momentum resolution.
This is particularly important for the ARPES pseudogap in electron-doped cuprates that has quite a detailed momentum space structure, and for d-wave superconducting correlations where the zero temperature pair correlation length may extend well beyond near-neighbor sites.
\item Although formulated as a lowest-order result of strong-coupling perturbation theory, it is not controlled by including higher-order terms in that perturbation expansion -- this would be extremely difficult -- but rather by increasing the cluster size.
\item It cannot describe broken-symmetry states.
This is accomplished by VCA and CDMFT, which can both be viewed as extensions or refinements of CPT.
\EE

\section{Exact Diagonalizations}\label{section:ED}

Before going on to describe more sophisticated quantum cluster approaches, let us describe in some detail a particular \textit{cluster solver}, i.e., a particular method used to calculate the ground state and Green function of the cluster: the exact diagonalization method, based on the Lanczos algorithm.
The quantum cluster methods described here are not tied to a specific solver for the cluster.
For instance, Quantum Monte Carlo or any other approximate method of solution for the cluster Green function could be used.
The exact diagonalization (ED) method has the advantage of high numerical accuracy at zero temperature, and can be to some extent controlled by the size of the cluster used.

The basic idea behind exact diagonalization is one of brute force, but its practical implementation may require a lot of care depending on the desired level of optimization.
Basically, an exact representation of the Hamiltonian action on arbitrary state vectors must be coded -- this may or may not involve an explicit construction of the Hamiltonian matrix.
Then the ground state is found in an quasi-exact way by an iterative method such as the Lanczos algorithm.
The Green function is thereafter calculated by similar means to be described below.
The main difficulty with execution is the large memory needed by the method, which grows exponentially with the number of degrees of freedom.
As for coding, the main difficulty is to optimize the method, in particular by taking point group symmetries into account.

\subsection{Coding of the basis states}

The first step in the exact diagonalization procedure is to define a coding scheme for the quantum basis states.
Let the different orbitals (or one-electron states) of the cluster be labelled by a greek index $\mu$, that is in fact of compound index of cluster position $\Rv$ and spin/band $\s$.
A basis state may be specified by the occupation number $n_{\mu}$ ($=0$ or 1) of electrons in the orbital labelled $\mu$ and has the following expression in terms of creation operators:
\BEQ
\label{eq:basis}
(c^\dg_{1\up})^{n_{1\up}}\cdots (c^\dg_{L\up})^{n_{L\up}}
(c^\dg_{1\dn})^{n_{1\dn}}\cdots (c^\dg_{L\dn})^{n_{L\dn}}|0\R
\EEQ
where the order in which the creation operators are applied is a matter of convention, but important.
If the number of orbitals is smaller than or equal to 32, the string of $n_\mu$'s forms the binary representation of a 32-bit unsigned integer $b$, which can be split into spin up and spin down parts:
\BEQ
b = b_\up + 2^{L} b_\dn
\EEQ

There are $2^{2L}$ such states, but not all are relevant, since the Hubbard Hamiltonian is block-diagonal~: The number of electrons of a given spin ($N_\up$ and $N_\dn$) is conserved and commutes with the Hamiltonian $H$.
Therefore the exact diagonalization is to be performed in a sector (i.e. a subspace) of the total Hilbert space with fixed values of $N_\up$ and $N_\dn$.
This space has the tensor product structure
\BEQ\label{eq:Hilbert1}
V = V_{N_\up} \otimes V_{N_\dn}
\EEQ
and has dimension $d = d(N_\up) d(N_\dn)$, where 
\BEQ
d(N_\s) = {L!\over N_\s!(L-N_\s)!}
\EEQ
is the dimension of each factor, i.e., the number of ways to distribute $N_\s$ electrons among $L$ sites.

Note that the ground state $|\Omega\R$ of the Hamiltonian generally belongs to the sector $N_\up=N_\dn$.
For a half-filled, zero spin system ($N_\up=N_\dn=L/2$), this translates into $d= (L!/(L/2)!^2)^2$, which behaves like $4^L/L$ for large $L$: 
The size of the eigenproblem grows exponentially with system size.
By contrast, the non-interacting problem can be solved only by concentrating on one-electron states.
For this reason, exact diagonalization of the Hubbard Hamiltonian is restricted to systems of the order of 16 sites or less.

In practice, a generic state vector is represented by an $d$-component array of double precision numbers.
In order to apply or construct the Hamiltonian acting on such vectors, we need a way to translate the label of a basis state (an integer $i$ from $0$ to $d-1$), into the binary representation (\ref{eq:basis}).
The way to do this depends on the level of complexity of the Hilbert space structure.
In the simple case (\ref{eq:Hilbert1}), one needs, for each spin, to build a two-way look-up table that tabulates the correspondence between consecutive integer labels and the binary representation of the spin up (resp. spin down) part of the basis state.
Thus, given a binary representation $(b_\up,b_\dn)$ of a basis state $|b\R = |b_\up\R|b_\dn\R$, one immediately finds integer labels $I_\up(b_
\up)$ and $I_\dn(b_\dn)$ and the label of the full basis state may be taken as
\BEQ
i = I_\up(b_\up) + d_{N\up} I_\dn(b_\dn)
\EEQ
On the other hand, given a label $i$, the corresponding labels of each spin part are
\BEQ
i_\up = \mod(i,d_{N\up}) \qquad i_\dn = i/d_{N\up}
\EEQ
where integer division (i.e. without fractional remainder) is used in the above expression.
The binary representation $b$ is recovered by inverse tables $B$ as
\BEQ
b_\up = B_\up(i_\up) \qquad b_\dn = B_\dn(i_\dn) 
\EEQ

The next step is to construct the Hamiltonian matrix.
The particular structure of the Hubbard model Hamiltonian brings a considerable simplification in the simple case studied here.
Indeed, the Hamiltonian has the form 
\BEQ
H = K_\up\otimes 1 + 1\otimes K_\dn + V_{\rm int.}
\EEQ
where $K_\up$ only acts on up electrons and $K_\dn$ on down electrons, and where the Coulomb repulsion term
$V_{\rm int.}$ is diagonal in the occupation number basis.
Thus, storing the Hamiltonian in memory is not a problem~: the diagonal $V_{\rm int.}$ is stored (an array of size $d$), and the kinetic energy $K_\s$ (a matrix having a small fraction of $d_\s^2$ elements) is stored in sparse form.
Constructing this matrix, formally expressed as 
\BEQ
K = \sum_{a,b} t_{ab} c^\dg_a c_b~,
\EEQ
needs some care with the signs.
Basically, two basis states $|b_\s\R$ and $|b'_\s\R$ are connected with this matrix if their binary representations differ at two positions $a$ and $b$.
The matrix element is then $(-1)^{M_{ab}} t_{ab}$, where $M_{ab}$ is the number of occupied sites between $a$ and $b$, i.e., assuming $a<b$,
\BEQ
M_{ab} = \sum_{c=a+1}^{b-1} n_c
\EEQ
For instance, the two states $(10010110)$ and $(10011100)$ with $L=8$ are connected with the matrix element $+t_{46}$, where the sites are numbered from 0 to $L-1$.

Calculating the Hubbard interaction is straighforward: a bit-wise \textsc{and} is applied to the up and down parts of a binary state ($b_\up\,\&\,b_\dn$ in C or \cpp) and the number of set bits of the result is the number of doubly occupied sites in that basis state.

\subsection{The Lanczos algorithm for the ground state}

Next, one must apply the exact diagonalization method per se, using the Lanczos algorithm.
Generally, the Lanczos method\cite{SIAM:Ruhe} is used when one needs the extreme eigenvalues of a matrix too large to be fully diagonalized (e.g. with the  Householder algorithm).
The method is iterative and involves only multiply-add's from the matrix.
This means in particular that the matrix does not necessarily have to be constructed explicitly, since only its action on a vector is needed.
In some extreme cases where it is practical to do so, the matrix elements can be calculated `on the fly', and this allows to save the memory associated with storing the matrix itself.

The basic idea behing the Lanczos method is to build a projection $\Hc$ of the full Hamiltonian matrix $H$ onto the so-called Krylov subspace.
Starting with a (random) state $|\phi_0\R$, the Krylov subspace is spanned by the iterated application of $H$:
\BEQ\label{eq:Krylov}
\mathscr{K} = \textrm{span}\l\{ |\phi_0\R, H|\phi_0\R, H^2|\phi_0\R, \cdots, H^M|\phi_0\R\r\}
\EEQ
the generating vectors above are not mutually orthogonal, but a sequence of mutually orthogonal vectors can be built from the following recursion relation
\BEQ\label{eq:LanczosRecursion}
|\phi_{n+1}\R = H|\phi_n\R -a_n|\phi_n\R -b_n^2|\phi_{n-1}\R
\EEQ
where 
\BEQ
a_n = {\L\phi_n|H|\phi_n\R\over \L\phi_n|\phi_n\R}
\qquad b_n^2 = {\L\phi_n|\phi_n\R\over \L\phi_{n-1}|\phi_{n-1}\R}
\qquad b_0 = 0
\EEQ
and we set the initial conditions $b_0=0$, $|\phi_{-1}\R=0$.
At any given step, only three state vectors are kept in memory ($\phi_{n+1}$, $\phi_n$ and $\phi_{n-1}$).
In the basis of normalized states $|n\R = |\phi_n\R/\sqrt{\L\phi_n|\phi_n\R}$, the projected Hamiltonian has the tridiagonal form 
\BEQ
\label{eq:tridiag}
H = \MATRIX{ 
a_0 & b_1 & 0 & 0 &\cdots & 0 \\ b_1 & a_1 & b_2 & 0 & \cdots & 0 \\
0  & b_2 & a_2 & b_3 & \cdots & 0 \\ \vdots & \vdots & \vdots & \vdots & \ddots & \vdots \\
0 & 0 & 0 & 0 & \cdots & a_N}
\EEQ
Such a matrix is readily diagonalized by fast methods dedicated to tridiagonal matrices, and a convergence criterion must be set for the lowest eigenvalue $E_0$, at which iterations stop.
For instance, one may stop the procedure when the lowest eigenvalue $E_0$ changes by no more that one part in $10^{12}$.
This may require between a number $M$ of iterations between a few tens and $\sim 200$, depending on system size.

The ground state energy $E_0$ and the ground state $|\Omega\R$ are very well approximated by the lowest eigenvalue and the corresponding eigenvector of that matrix, which are obtained by standard methods.
This provides us with the ground state $|\Omega\R$ in the reduced basis $\{|\phi_n\R\}$. But we need the ground state in the original basis, and this requires retracing the Lanczos iterations a second time -- for the $|\phi_n\R$ are not stored in memory -- and constructing the ground state progressively at each iteration from the known coefficients $\L\Omega|\phi_n\R$.

The Lanczos procedure is simple and efficient.
Convergence is fast if the lowest eigenvalue $E_0$ is well separated from the next one ($E_1$).
It slows down if $E_1-E_0$ is small.
If the ground state is degenerate ($E_1=E_0$), the procedure will converge to a vector of the ground state subspace, a different one each time the initial state $|\phi_0\R$ is changed.

Note that the sequence of Lanczos vectors $|\phi_n\R$ is in principle orthogonal, as this is garanteed by the three-way recursion relation (\ref{eq:LanczosRecursion}).
However, numerical error will introduce `orthogonality leaks', and after a few tens of iterations the Lanczos basis will become overcomplete in the Krylov subspace.
This will translate in multiple copies of the ground state eigenvalue in the tridiagonal matrix (\ref{eq:tridiag}), which should not be taken as a true degeneracy.
However, as long as one is only interested in the ground state and not in the multiplicity of the lowest eigenvalues, this is not a problem.

\subsection{The Lanczos algorithm for the Green function}

Once the ground state is known, it remains to calculate the cluster Green function.
The zero-temperature Green function $G_{\mu\nu}(\om)$ has the following expression, as a function of the complex-valued frequency $\om$:
\BEA\label{eq:green}
G'_{\mu\nu}(\om) & = & G'_{\mu\nu,e}(\om) + G'_{\mu\nu,h}(\om) \\
G'_{\mu\nu,e}(\om) & = & \L\Omega|c_\mu\frac1{\om-H+E_0}c^\dg_\nu|\Omega\R \\
G'_{\mu\nu,h}(\om) & = & \L\Omega|c^\dg_\nu\frac1{\om+H-E_0}c_\mu|\Omega\R 
\EEA
In the basic Hubbard model, spin is conserved and we need only to consider the creation and annihilation of up-spin electrons.

We will first describe a Lanczos algorithm for calculating the Green function, that provides a continued-fraction representation of its frequency dependence.
In the next subsection, we will instead present an alternate method based on the Band Lanczos algorithm, that provides a Lehmann representation of the Green function and that is both faster and more memory intensive.

Consider first the function $G'_{\mu\mu,e}(\om)$.
One needs to know the action of $(\om-H+E_0)^{-1}$ on the state $|\phi_\mu\R = c_\mu^\dg|\Omega\R$, and then to calculate
\BEQ
\label{eq:GreenCF1}
G'_{\mu\mu,e} = \L\phi_\mu|\frac1{\om - H + E_0}|\phi_\mu\R
\EEQ
As with any generic function of $H$, this one can be expanded in powers of $H$:
\BEQ
\frac1{z-H} = \frac1{z} + \frac1{z^2}H + \frac1{z^3}H^2 + \cdots
\EEQ
and the action of this operator can be evaluated exactly at order $H^M$ in a Krylov subspace (\ref{eq:Krylov}).
Thus we again resort to the Lanczos algorithm: A Lanczos sequence is calculated from the initial, normalized state $|\phi_0\R=|\phi_\mu\R/b_0$ where $b_0^2 = \L\phi_\mu|\phi_\mu\R$.
This sequence generates a tridiagonal representation of $H$, albeit in a different Hilbert space sector~: that with $N_\up+1$ up-spin electrons and $N_\dn$ down-spin electrons.
Once the preset maximum number of Lanczos steps, or a near zero value of $b_n$, has been reached, the tridiagonal representation (\ref{eq:tridiag}) may then be used to calculate (\ref{eq:GreenCF1}).
This amounts to the matrix element $b_0^2[(\om - H + E_0)^{-1}]_{00}$ (the first element of the inverse of a tridiagonal matrix), which has a simple continued fraction form~:\cite{Dagotto:1994vn}
\BEQ
\label{eq:jacobi}
G'_{\mu\mu,e}(\om) = \cfrac{b_0^2}{\om - a_0 - \cfrac{b_1^2}{\om - a_1 - \cfrac{b_2^2}{\om - a_2 -\cdots}}}
\EEQ
Thus, evaluating the Green function, once the arrays $\{a_n\}$ and $\{b_n\}$ have been found, reduces to the calculation of a truncated continued fraction, which can be done recursively in $M$ steps, starting from the bottom floor of the fraction.

Consider next the case $\mu\ne\nu$.
The continued fraction representation applies only to the case where the same state $|\phi\R$ appears on the two sides of (\ref{eq:GreenCF1}).
If $\mu\ne\nu$, this is no longer the case, but we may use the following trick~:
we define the combination 
\BEQ
G_{\mu\nu,e}^+(\om) = \L\Omega|(c_\mu + c_\nu)\frac1{\om-H+E_0}(c_\mu + c_\nu)^\dg|\Omega\R
\EEQ
Using the symmetry $G_{\mu\nu,e}(\om) = G_{\nu\mu,e}(\om)$, this leads to 
\BEQ
G_{\mu\nu,e}(\om) = \frac12 (G^+_{\mu\nu,e}(\om) - G_{\mu\mu,e}(\om) - G_{\nu\nu,e}(\om))
\EEQ
where $G_{\mu\nu,e}^+$ can be calculated in the same way as $G_{\mu\mu,e}$, i.e., with a simple continued fraction.
We proceed likewise for $G_{\mu\nu,h}^+(\om)$.

Thus, the cluster Green function is encoded in $L(L+1)$ continued fractions, whose coefficients are stored in memory, so that $\Gv'(\om)$ can be computed on demand for any complex frequency $\om$.

Note that a minimal way to take advantage of cluster symmetries is to restrict the calculation of the Green function to an irreducible set of pairs $(\mu,\nu)$ of orbitals that can generate all other pairs by symmetry operations of the cluster.
Thus, if a symmetry operation $g$ takes the orbital $\mu$ into the orbital $g(\mu)$, we have
\BEQ
G'_{\mu\nu}(\om) = G'_{g(\mu)g(\nu)}(\om)
\EEQ
Taking this into account is an easy and important time saver, but not as efficient as using a basis of symmetry eigenstates, as described later on in this section. 

\subsection{The Band Lanczos algorithm for the Green function}

An alternate way of calculating the cluster Green function is to apply the \textit{band Lanczos} procedure\cite{SIAM:Freund}.
This is a generalization of the Lanczos procedure in which the Krylov subspace is spanned not by one, but by many states.
Let us assume that up and down spins are decoupled, so that the Green function is $L\times L$ block diagonal.
The $L$ states $|\phi_\mu\R = c^\dg_\mu|\Omega\R$ are first constructed, and then one builds the projection $\Hc$ of $H'$ on the Krylov subspace spanned by
\BEQ\begin{split}
\Big\{ |\phi_1\R,\dots,|\phi_{L}\R,&H'|\phi_1\R,\dots,H'|\phi_{L}\R,\dots,\\
&(H')^M|\phi_1\R,\dots,(H')^M|\phi_{L}\R\Big\}
\end{split}
\EEQ
A Lanczos basis $\{|n\R\}$ is constructed by successive application of $H'$ and orthonormalization with respect to the previous $2L$ basis vectors.
In principle, each new basis vector $|n\R$ is already automatically orthogonal to basis vectors $|1\R$ through $|n-2L-1\R$, although `orthogonality leaks' arise eventually and may be problematic.
A practical rule of thumb to avoid these problems is to control the number $M$ of iterations by the convergence of the lowest eigenvalue of $
\Hc$ (e.g. to one part in $10^{10}$).
Independently of this, one must be careful about potential redundant basis vectors in the Krylov subspace, which must be properly `deflated'.
\cite{SIAM:Freund}
The number of states $R$ in the Krylov subspace at convergence is typically between 100 and 300, depending on system size.
The $R\times R$ matrix $\Hc$, which has a tridiagonal structure in the ordinary Lanczos method, now has a band structure made of $2L$ diagonals around the central diagonal.
It is then a simple matter to obtain a Lehmann representation of the Green function in the Krylov subspace (see Appendix~\ref{section:Lehmann}) by calculating the projections $Q_{\mu r}$ of $|\phi_\mu\R$ on the eigenstates of $\Hc$ (the inner products of the $|\phi_\mu\R$'s with the Lanczos vectors are calculated as the latter are constructed).
The Green function can then be expressed in a Lehmann representation (\ref{eq:qmatrix1}).
The two contributions $G'_{\mu\nu,e}$ and $G'_{\mu\nu,h}$ to the Green function are computed separately, and the corresponding matrices $\Qv$ and $\Lambdav$ are simply concatenated to form the complete $\Qv$- and $\Lambdav$-matrices, which are then stored and allow again for a quick calculation of the Green function as a function of the complex frequency $\om$.
The matrix $2L\times R$ matrix $\Qv$ has the property that
\BEQ
\Qv \Qv^\dg = \mathsf{1}_{2L\times 2L} 
\EEQ
This holds even if the Lehmann representation is obtained from a subspace and not the full space, and is simply a consequence of the anticommutation relations $\{c_\mu,c_\nu^\dg\} = \delta_{\mu\nu}$.

The band Lanczos method requires more memory than the usual Lanczos method, since $2L+1$ vectors must simultaneously be kept in memory, compared to $3$ for the simple Lanczos method.
On the other hand, it is faster since all pairs $(\mu,\nu)$ are covered in a single procedure, compared to $L(L+1)/2$.
Thus, we gain a factor $L^2$ in speed at the cost of a factor $L$ in memory.
Another advantage is that it provides a Lehmann representation of the Green function.

\subsection{Cluster symmetries}
\label{subsec:symmetries}

It is possible to optimize the exact diagonalization procedure by taking advantage of the symmetries of the cluster Hamiltonian, in particular coming from cluster geometry.
If the Hamiltonian is invariant under a discrete group $\Gg$ of symmetry operations and $|\Gg|$ denotes the number of such elements (the order of the group), the dimension of the largest Hilbert space needed can be reduced by a factor of almost $|\Gg|$, and the number of state vectors needed in the band Lanczos method reduced by the same factor.
The corresponding speed gain is appreciable.
In the case of large clusters (e.g. 16 sites), taking advantage of symmetries may make the difference between doing or not doing the problem.
The price to pay is a higher complexity in coding the basis states, which almost forces one to store the Hamiltonian matrix in memory, if it were not already, since calculating matrix elements `on the fly' becomes more time consuming.
Note that we are using open boundary conditions (except in the case of the DCA, not discussed in these notes), and therefore there is no translation symmetry within the cluster; thus we are concerned with points groups, not space groups.

Let us start with a simple example: a cluster invariant with respect to a single inversion, or a single rotation by $\pi$.
One may think of a one-dimensional cluster, for instance, with a left-right inversion.
The corresponding symmetry group is $C_2$, with two elements: the identity $e$ and the inversion $\iota$.
The group $C_2$ contains two irreducible representations, noted $A$ and $B$, corresponding respectively to states that are even and odd with respect to $\iota$.
Because the Hamiltonian is invariant under inversion: $H = \iota^{-1}H\iota$, eigenvectors of $H$ will be either even or odd, i.e. belong either to the A or to the B representation.
Likewise, the Hamiltonian will have no matrix elements between states belonging to different representations (the reader is invited to read Appendix \ref{section:groups} for a review of the necessary group-theoretical concepts).

In order to take advantage of this fact, one needs to construct a basis containing only states of a given representation.
The occupation number basis states $|b\R$ (or binary states, as we will call them) introduced above are no longer adequate.
In the case of the simple group $C_2$, one should rather consider the even and odd combinations $|b\R \pm \iota|b\R$ (and some of these combinations may vanish).
Yet we still need a scheme to label the different basis states and have a quick access to their occupation number representation, which allows us to compute matrix elements.
Let us briefly describe how this can be done (a more detailed discussion can be found, e.g., in Ref.~\onlinecite{Poilblanc:2004}).
Under the action of the group $\Gg$, each binary state generates an `orbit' of binary states, whose length is the order $|\Gg|$ of the group, or a divisor thereof.
To such an orbit corresponds at most $d_\a$ states in the irreducible representation labeled $\a$, given by the corresponding projection operator:
\BEQ
|\psi\R = \frac{d_\a}{|\Gg|}\sum_g \chi^{(\a)*}_g g|b\R
\EEQ
where $d_\a$ is the dimension of the irreducible representation $\a$.
We will restrict the discussion to the simplest case, where all irreducible representations considered are one-dimensional ($d_\a=1$; the case $d_\a > 1$ turns out to be quite a bit more complex).
Then the state $|\psi\R$ is either zero or unique for a given orbit.
We can then select a representative binary state for each orbit (e.g. the one associated with the smallest binary representation) and use it as a label for the state $|\psi\R$.
We still need an index function $B(i)$ which provides the representative binary state for each consecutive label $i$.
The reverse correspondence $i=I(b)$ is trickier, since symmetrized states are no longer factorized as products of up and down spin parts.
It is better then to search the array $B$ for the value of the index $i$ that provides a given binary state $b$.
One can still be aided by a partial reverse index $I_\up(b_\up)$ that provides the first occurence in the list $B$ of a state with $b_\up$ as the spin up part, assuming that states are sorted according to $b_\up$, then according to $b_\dn$.

Once the basis has been constructed, one needs to construct a matrix representation of the Hamiltonian in that representation.
Given two states $|\psi_1\R$ and $|\psi_2\R$, represented by the binary states $|b_1\R$ and $|b_2\R$, it is a simple matter to show that the matrix element is
\BEQ\label{eq:matrixElement}
\L\psi_2 | H | \psi_1\R = \frac{d_\a}{|\Gg|}\sum_g \chi_h^{(\a)*} \phi_g(b) \L gb_2|H|b_1\R
\EEQ
where the phase $\phi_g(b)$ is defined by the relation
\BEQ
g|b\R = \phi_g(b)|gb\R~.
\EEQ
In the above relation, $|gb\R$ is the binary state obtained by applying the symmetry operation $g$ to the occupation numbers forming $b$, whereas the phase $\phi_g(b)$ is the product of signs collected from all the permutations of creation operators needed to go from $b$ to $gb$.
Formula (\ref{eq:matrixElement}) is used as follows to construct the Hamiltonian matrix:
First, the Hamiltonian can be written as $H = \sum_r H_r$, where $H_r$ is a hopping term between specific sites, or a diagonal term like the interaction.
One then loops over all $b_1$'s.
For each $b_1$, and each term $H_r$, one construct the single binary state $H_r|b_1\R$.
One then finds the representative $b_2$ of that binary state, by applying on it all possible symmetry operations until $g$ is found such that $|gb_2\R =  H_r|b_1\R$.
During this operation, the phase $\phi_g(b)$ must also be collected.
Then the matrix element (\ref{eq:matrixElement}) is added to the list of stored matrix elements.
Since each term $H_r$ individually is not invariant under the group, there will be more matrix elements generated than there should be, i.e., there will be cancellations between different matrix elements associated with the same pair ($b_1$, $b_2$) and produced by the different $H_r$'s.
For this reason, it is useful to first store all matrix elements associated with a given $b_1$ in an intermediate location in order for the cancellations to take effect, and then to store the cleaned up `column' labelled by $b_1$ to its definitive storage location.
Needless to say, one should only store the row and column indices of each element of a given value.

\begin{table}
\centering
\caption{Number of matrix elements of a given value in the nearest-neighbor hopping operator on the half-filled $3\times4 = 12$ site cluster, for each irreducible representation of $C_{2v}$. The dimension of each subspace is indicated on the second row.\label{table:3x4stats}}
\begin{tabular}{R|RRRR}
\hline
& A_1 & A_2 & B_1 & B_2 \\
\mathrm{dim.} & 213,840 & 213,248 & 213,440 & 213,248 \\
\mathrm{value} & & & & \\
\hline
-2	& 96	& 736	& 704	& 0 \\
-\sqrt{2}	& 12,640	& 6,208	& 7,584	& 5,072 \\
-1	& ~~2,983,264	& ~~2,936,144	& ~~2,884,832	& ~~2,911,920 \\
1	& 952,000	& 997,168	& 1,050,432	& 1,021,392 \\
\sqrt{2}	& 5,088	& 2,304	& 3,232	& 2,992 \\
2	& 32	& 0	& 0	& 0 \\
\hline
\end{tabular}
\end{table}

Table \ref{table:3x4stats} gives the values and number of matrix elements found for the nearest-neighbor hopping terms on the half-filled 12-site ($3\times 4$) cluster, in each of the four irreducibe representations of the group $C_{2v}$.

\subsection{Green functions using cluster symmetries}

Most of the time, the ground state lies in the trivial (symmetric) representation.
However, taking advantage of symmetries in the calculation of the Green function requires all the irreducible representations to be included in the calculation.
Consider for instance the simple example of a $C_2$ symmetry, with a ground state $|\Omega\R$ in the $A$ (even) representation.
Constructing the Green function involves applying on $|\Omega\R$ the destruction operator $c_a$ (or the creation operator $c^\dg_a$) associated to site $a$.
The excited state thus produced does not belong to a well-defined representation.
Instead, on should destroy (or create) and electron in an odd or even state, by using the linear combinations $c_a \pm c_{\iota a}$, where $\iota a$ is the site obtained by applying the symmetry operation to $a$.
Thus, in calculating the Green function (\ref{eq:green}), one should express each creation/destruction operator in terms of symmetrized combinations, e.g., 
\BEQ
c_a = \hf(c_a + c_{\iota a}) + \hf(c_a - c_{\iota a})
\EEQ
More generally, one would use symmetrized combinations of operators
\BEQ
c^{(\a)}_\rho = \sum_a M^{(\a)}_{\rho a}c_a
\EEQ
such that $c^{(\a)}_\rho$ transforms under representation $\a$, and $\rho$ labels the different possibilities.
For instance, for a linear cluster of length 4 and an inversion symmetry that maps the sites $(1234)$ into $(4321)$, these operators are
\BEQ
\ALIGNED{c^{(A)}_1 &= c_1 + c_4 \\ c^{(A)}_2 &= c_2 + c_3} \quad
\ALIGNED{c^{(B)}_1 &= c_1 - c_4 \\ c^{(B)}_2 &= c_2 - c_3}
\EEQ
Then, for each representation, one may use the Band Lanczos procedure and obtain a Lehmann representation $Q^{(\a)}_{\rho r}$ for the associated Green function $G_{\rho\s}^{(\a)}(\om)$.
If the ground state is in representation $\a$ and the operators $c^{(\b)}_\rho$ of representation $\b$ are used, the Hilbert space sector to work with will be the tensor product representation $\a\otimes\b$, which poses no problem at all when all irreps are one-dimensional, but would bring additional complexity if the ground state were in a multi-dimensional representation.
Finally, one may bring together the different pieces, by building a $L\times L$ matrix $M_{\rho a}$ that is the vertical concatenation of the various rectangular matrices $M^{(\a)}_{\rho a}$, and returning to the usual $\Qv$-matrix representation
\BEQ
Q_{ar} = (\Mv^{-1})_{a\rho} Q_{\rho r}
\EEQ
Using cluster symmetries for the Green function saves a factor $|\Gg|$ in memory because of the reduction of the Hilbert space dimension, and an additional factor of $|\Gg|$ since the number of input vectors in the band Lanczos procedure is also divided by $|\Gg|$.
Typically then, most of the memory will be used to store the Hamiltonian matrix.

\subsection{Parallelization}

For larger clusters (e.g. 16 sites), the computer memory required to carry out the exact diagonalization is too large to fit on a typical computer.
In those cases the only practical choice is to parallelize the exact diagonalization procedure.
Although this is a technical issue that has more to do with programming than with the algorithm, a brief explanation is in order.
Parallelization consists in dividing the task and data between many processes (run on different cpus), with communication between processes taking place on a frequent basis.
The Message Passing Interface (MPI) Library is the most common way to accomplish this on distributed-memory machines.
Parallelization is often a difficult task, and is likely not to scale well (i.e., the inverse computing time grows more slowly than the number of processes) when inter-process communications occur too frequently.
However, parallelization makes the difference between doing or not doing a large problem.

\InsertFigure{How to split a matrix-vector multiplication $|y\R=H|x\R$ across two processes.
`Blue' data reside on one process, and `red' on the other.
For each of the two segments of $|y\R$, each process performs a block matrix multiplication, and the results of the two processes must be transfered to each other to be added.\label{fig:parallel}}{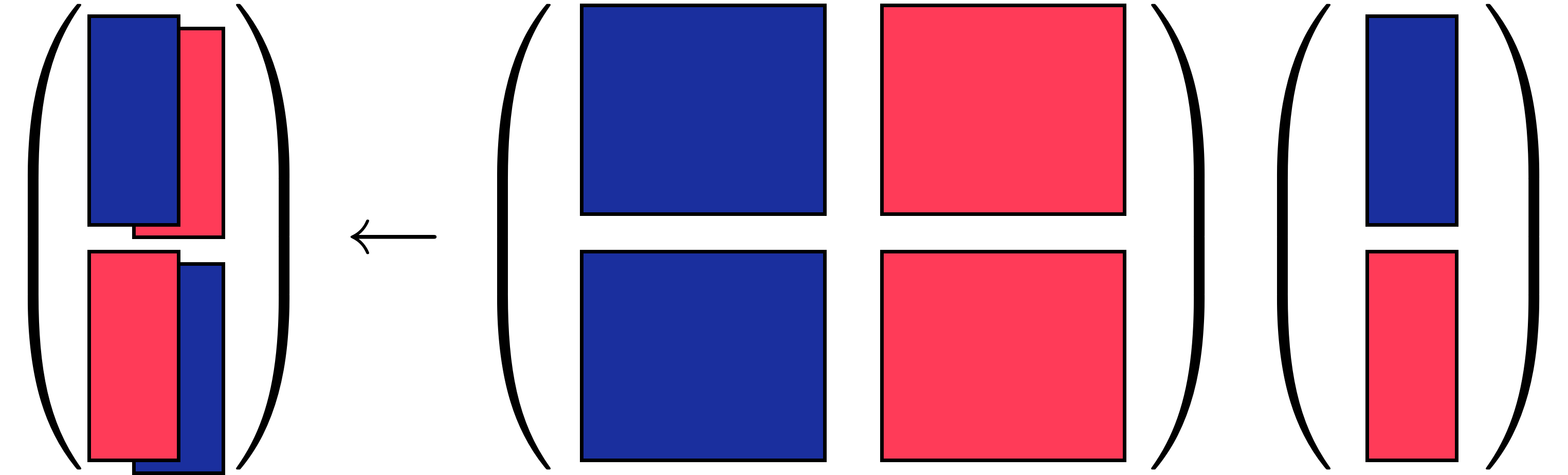}{0.8\hsize}

Let us now briefly describe a possible way to parallelized an exact diagonalization program, as used by us.
Let $\Nc$ be the number of processes across which the problem is parallelized.
We split each Hilbert space vector into $\Nc$ (nearly) equal segments, and the Hamiltonian matrix into $\Nc^2$ blocks (labelled $H_{IJ}$, with $I,J=1,\dots,\Nc$.
A single matrix-vector multiplication $|y\R=H|x\R$ then proceeds, for each process, by $\Nc$ successive operations $|y\R_I = H_{IJ}|x\R_J$, $J$ labelling the different processes and $I$ the successive operations.
After each operation the resulting vectors $|y\R_I$ must be sent to process $I$ to be summed in a single segment.
This is illustrated on Fig.~\ref{fig:parallel} for $\Nc=2$.
Thus, each multiply-add operator involves $\Nc$ `broadcast' or `reduce' operations, in MPI jargon.
The construction of the Hamiltonian is also parallelized, as each process takes care of its own group of columns.
This constitutes what is called \textit{fine-grained} parallelization: communications are very frequent (many calls per matrix-vector multiply add).
Consequently, scaling is poor and in practice the number of processes should be kept to a minimum, just enough to fit the program in memory.

As a whole, computational scientists will feel an ever increasing pressure to use parallel computing, as this will become the only way not only to do larger problem, but to substantially speed up all problems, because of the slowing down of Moore's law and of the ubiquity of cpus with an increasing number of cores.

\section{The self-energy functional approach}

That CPT is incapable of describing broken symmetries is its major drawback.
Treating spontaneously broken symmetries requires some sort of self-consistent procedure, or a variational principle.
Ordinary mean-field theory does precisely that, but is limited by its discarding of fluctuations and its uncontrolled character.

A heuristic way of treating broken symmetry states within CPT would be to add to the cluster Hamiltonian $H'$ a Weiss field that pushes the system towards some predetermined form of order.
For instance, the following term, added to the Hamiltonian, would induce N\'eel antiferromagnetism:
\BEQ\label{eq:weissAF}
H'_M = M\Oc_M \equiv M\sum_\Rv \er^{i\Qv\cdot\Rv}(n_{\Rv\up}-n_{\Rv\dn})
\EEQ
where $\Qv=(\pi,\pi)$ is the antiferromagnetic wavevector.
What is needed is a procedure to set the value of the Weiss parameter $M$.
Adopting a mean-field-like procedure (i.e. factorizing the interaction in the correct channel and applying a self-consistency condition) would bring us exactly back to ordinary mean-field theory: the interaction having disappeared, the cluster decomposition would be suddenly useless and CPT would provide the same result regardless of cluster size.

The solution to that conundrum is most elegantly provided by the self-energy functional approach (SFA), proposed by Potthoff.\cite{Potthoff:2003}
This approach also has the merit of presenting various cluster schemes from a unified point of view.
It can also be seen as a special case of the more general inversion method\cite{Fukuda:1995}, recently reviewed in Ref.~\onlinecite{Kotliar:2006kx} in the context of Density Functional Theory and DMFT.

To start with, let us introduce a functional $\Omega_\tv[\Gv]$ of the Green function:
\BEQ\label{eq:GrandPotential}
\Omega_\tv[\Gv]=\Phi[\Gv]-\Tr((\Gv_{0\tv}^{-1} -\Gv^{-1})\Gv)+\Tr\ln(-\Gv).
\EEQ
This means that, given any Green function $G_{ij}(\om)$ one can cook up -- yet with the usual analytic properties of Green functions as a function of frequency -- this expression yields a number.
In the above expression, products and powers of Green functions -- e.g. in series expansions like that of the logarithm -- are to be understood in a functional matrix sense.
This means that position $i$ and time $\tau$, or equivalently, position and frequency, are merged into a single index.
Accordingly, the symbol $\Tr$ denotes a functional trace, i.e., it involves not only a sum over sites indices, but also over frequencies.
The latter can be taken as a sum over Matsubara frequencies at finite temperature, or as an integral over the imaginary frequency axis at zero temperature.

\InsertFigure{Diagrammatic definition of the Luttinger-Ward functional, as a sum over two-particle irreducible graphs.\label{fig:skeleton}}{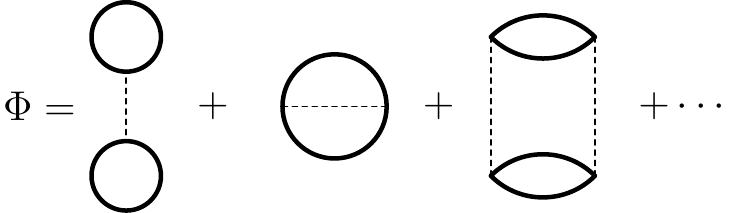}{0.8\hsize}

The Luttinger Ward functional $\Phi[\Gv]$ entering this expression is usually defined as the sum of \textit{two-particle irreducible} (2PI) diagrams : diagrams that cannot by split into disjoint parts by cutting two fermion lines (Fig.~\ref{fig:skeleton}).
These are sometimes called \textit{skeleton} diagrams, although `two-particle irreducible' is more accurate.
A diagram-free definition of $\Phi[\Gv]$ is also given in Ref.~\onlinecite{Potthoff:2006os}.
For our purposes, what is important is that (1) The functional derivative of $\Phi[\Gv]$ is the self-energy 
\BEQ\label{eq:SelfLuttinger}
\frac{\delta\Phi[\Gv]}{\delta \Gv}=\Sigmav
\EEQ
(as defined diagramatically) and (2) it is a universal functional of $\Gv$ in the following sense: whatever the form of the one-body Hamiltonian, it depends only on the interaction and, functionnally, it has the same dependence on $\Gv$.
This is manifest from its diagrammatic definition, since only the interaction (dotted lines) and the Green function given as argument, enter the expression.
The dependence of the functional $\Omega_\tv[\Gv]$ on the one-body part of the Hamiltonian is denoted by the subscript $\tv$ and it comes only through $\Gv_{0\tv}^{-1} = \om-\tv$ appearing on the right-hand side of Eq.~(\ref{eq:GrandPotential}).

The functional $\Omega_\tv[\Gv]$ has the important property that it is stationary when $\Gv$ takes the value prescribed by Dyson's equation.
Indeed, given the last two equations, the Euler equation takes the form
\BEQ
\frac{\delta\Omega_\tv[\Gv]}{\delta \Gv}=\Sigmav-\Gv_{0\tv}^{-1}+\Gv^{-1}=0.
\EEQ
This is a dynamic variational principle since it involves the frequency appearing in the Green function, in other words excited states are involved in the variation.
At this stationary point, and only there, $\Omega_\tv[\Gv]$ is equal to the physical (thermodynamic) grand potential.
Contrary to Ritz's variational principle, this last equation does not tell us whether $\Omega_\tv[\Gv]$ is a minimum, a maximum, or a saddle point there.

There are various ways to use the stationarity property that we described above.
The most common one is to approximate $\Phi[\Gv]$ by a finite set of diagrams.
This is how one obtains the Hartree-Fock, the FLEX
approximation\cite{Bickers:1989} or other so-called thermodynamically consistent theories.
This is what Potthoff calls a type II
approximation strategy.\cite{Potthoff:2005} A type I approximation simplifies the Euler equation itself.
In a type III approximation, one uses the exact form of $\Phi[\Gv]$ but only on a limited domain of trial Green functions.

Following Potthoff, we adopt the type III approximation on a functional of the self-energy instead of on a functional of the Green function.
Suppose we can locally invert Eq.~(\ref{eq:SelfLuttinger}) for the self-energy  to write $\Gv$ as a functional of $\Sigmav$.
We can use this result to write,
\BEQ\label{eq:sef1}
\Omega_\tv[\Sigmav]=F[\Sigmav]-\Tr\ln(-\Gv_{0\tv}^{-1}+\Sigmav).
\EEQ
where we defined
\BEQ
F[\Sigmav]=\Phi[\Gv]-\Tr(\Sigmav\Gv).
\EEQ
and where it is implicit that $\Gv=\Gv[\Sigmav]$ is now a functional of $\Sigmav$.
$F[\Sigmav],$ along with the expression (\ref{eq:SelfLuttinger}) for the derivative of the Luttinger-Ward functional, defines the Legendre transform of the Luttinger-Ward functional.
It is easy to verify that
\BEQ
\varia{F[\Sigmav]}{\Sigmav}=\varia{\Phi[\Gv]}{\Gv}\varia{\Gv[\Sigmav]}{\Sigmav}-\Sigmav\varia{\Gv[\Sigmav]}{\Sigmav}-\Gv=-\Gv
\EEQ
hence, $\Omega_\tv[\Sigmav]$ is stationary with respect to $\Sigmav$
when Dyson's equation is satisfied
\BEQ
\varia{\Omega_\tv[\Sigmav]}{\Sigmav}=-\Gv+(\Gv_{0\tv}^{-1}-\Sigmav)^{-1}=0.
\EEQ

To perform a type III approximation on $F[\Sigmav]$, we take advantage that it is universal, i.e., that it depends only on the interaction part of the Hamiltonian and not on the one-body part.
We then consider another Hamiltonian, denoted $H'$ and called the \textit{reference system}, that describes the same degrees of freedom as $H$ and shares the same interaction (i.e. two-body) part.
Thus $H$ and $H'$ differ only by one-body terms.
We have in mind for $H'$ the cluster Hamiltonian, or rather the sum of all (mutually decoupled) cluster Hamiltonians.
At the physical self-energy $\Sigmav$ of the cluster, Eq.~(\ref{eq:sef1}) allows us to write
\BEQ
\Omega_{\tv'}[\Sigmav] = \Omega' = F[\Sigmav]-\Tr\ln(-\Gv')~,
\EEQ
where $\Omega'$ is the cluster Hamiltonian's grand potential and $\Gv'$ its physical Green function, obtained through the exact solution.
From this we can extract $F[\Sigmav]$ and it follows that
\BEQ\label{eq:sef}
\ALIGNED{
\Omega_\tv[\Sigmav] &= \Omega'+\Tr\ln(-\Gv')-\Tr\ln(-\Gv_{0\tv}^{-1}+\Sigmav) \\
&= \Omega'+\Tr\ln(-\Gv')-\Tr\ln(-\Gv)
}\EEQ
where $\Gv$ now stands for the CPT Green function (\ref{eq:CPT0}).
This expression can be further simplified as
\BEQ\label{eq:sef2}
\Omega_\tv[\Sigmav] = \Omega' - \Tr\ln(\id-\Vv\Gv') 
\EEQ
Let us finally make the trace more explicit: It is a sum over frequencies and a sum over lattice sites (and spin and band indices), which can be expressed instead as a sum over reduced wavevectors (as the CPT Green function is diagonal in that index), plus a ``small'' trace (denoted $
\tr$) on residual indices (cluster site, spin, and band):
\ALIGN{\label{eq:sef3}
\Omega_\tv[\Sigmav] &= \Omega' - T\sum_{\om}\sum_\kvt \tr\ln\l[1-\Vv(\kvt)\Gv'(\om)\r] \notag\\
&= \Omega' - T\sum_\om\sum_\kvt \ln\det\l[\id-\Vv(\kvt)\Gv'(\om)\r]
}
where the matrix identity $\tr\ln A = \ln\det A$ was used in the second equation.

The type III approximation comes from the fact that the self-energy $\Sigmav$ is restricted to the exact self-energy of the cluster problem $H'$, so that variational parameters appear in the definition of the one-body part of $H'$.
To come back to the question of the Weiss field $M$ introduced at the beginning of this section, we would set its value by solving the cluster Hamiltonian -- i.e., calculating $\Omega'$ and $\Gv'$ -- for many different values of $M$  and evaluate the functional (\ref{eq:sef3}) for each of them, selecting the value that makes Expression (\ref{eq:sef3}) stationary.
This is the idea behind the variational cluster approximation (VCA), described in more detail in the next section.

In practice, we look for values of the cluster one-body parameters
$\tv'$ such that $\delta\Omega_\tv[\Sigmav]/\delta\tv'=0$.
It is useful for what follows to write the latter equation formally, although we do not use it in actual calculations.
Given that $\Omega'$ is the actual grand potential evaluated for the cluster, $\partial\Omega'/\partial\tv'$ is canceled by the explicit $\tv'$ dependence of $\Tr\ln(-\Gv_{0\tv'}^{-1}+\Sigmav)$ and we are left with
\ALIGN{
0  &  =\varia{\Omega_\tv[\Sigmav]}{\Sigmav}\varia{\Sigmav}{\tv'} \notag\\
&  =-\Tr\l[\l(  \frac{1}{\Gv_{0\tv'}^{-1}-\Sigmav}-\frac{1}{\Gv_{0\tv}^{-1}-\Sigmav}\r)
\varia{\Sigmav}{\tv'}\r]~.
}
This may be explicited as
\ALIGN{\label{eq:EulerVCA}
&\sum_\om\sum_{\mu\nu}\Bigg[\l( \frac{1}{\Gv_{0\tv'}^{-1}-\Sigmav(\om)}\r)_{\mu\nu} \notag\\
& -\frac LN\sum_{\kvt}\l(  \frac{1}{\Gv_{0\tv}^{-1}(\kvt)-\Sigmav(\om)}\r)_{\mu\nu}\Bigg]\varia{\Sigma_{\nu\mu}'(\om)}{\tv'}=0.
}
where Greek indices are used for compound indices gathering cluster site, spin and possible band indices.

\section{The Variational Cluster Approximation}

The Variational Cluster Approximation\cite{Potthoff:2003,Dahnken:2004} (VCA), also called Variational Cluster Perturbation Theory (VCPT), can be viewed as an extension of Cluster Perturbation Theory in which some parameters of the cluster Hamiltonian are set according to Potthoff's variational principle through a search for saddle points of the functional (\ref{eq:sef3}).
The cluster Hamiltonian $H'$ is typically augmented by Weiss fields, such as the N\'eel field (\ref{eq:weissAF}) that allow for broken symmetries that would otherwise be impossible within a finite cluster.
The hopping terms and chemical potential within $H'$ may also be treated like additional variational parameters.
In contrast with Mean-Field theory, these Weiss fields are not mean fields, in the sense that they do not coincide with the corresponding order parameters.
The interaction part of $H$ (or $H'$) is not factorized in any way and short-range correlations are treated exactly.
In fact, the Hamiltonian $H$ is not altered in any way; the Weiss fields are introduced to let the variational principle act on a space of self-energies that includes the possibility of specific long-range orders, without imposing those orders.

Steps towards a VCA calculation are as follows:
\BE
\item Choose the Weiss fields to add, aided by intuition about the possible broken symmetries to expect.
\item Set up a procedure to calculate the functional (\ref{eq:sef3}).
\item Set up a procedure to optimize the functional, i.e., to find its saddle points, in the space of variational parameters.
\item Calculate the properties of the model the saddle point.
\EE

\subsection{Practical calculation of the Potthoff functional}

Let $\xiv$ denote the (finite) set of variational parameters to be used.
The Potthoff functional becomes the function
\BEQ\label{eq:sef4}
\Omega_\tv(\xiv) = \Omega' - \frac{TL}{N}\sum_\om\sum_\kvt \ln\det\l[\id-\Vv(\kvt)\Gv'(\kvt,\om)\r]
\EEQ
Once the cluster Green function is known by the methods described in Sect.~\ref{section:ED}, calculating the functional (\ref{eq:sef4}) requires an integral over frequencies and wavevectors of an expression that requires a few linear-algebraic operations to evaluate.
Two different methods have been used to compute these sums, described in what follows.
We will see that the second method, entirely numerical, is much faster than the first one, which is partly analytic, a result that may seem paradoxical.

\subsubsection{Method I : Exact frequency integration}\label{subsec:AI}

The integral over frequencies in (\ref{eq:sef3}) may be done analytically, with the result \cite{Aichhorn:2006rt}
\BEQ\label{eq:omega4}
\Om(\xiv) = \Omega'(\xiv) - \sum_{\om'_r<0} \om'_r + \frac LN\sum_\kvt \sum_{\om_r(\kvt)<0} \om_r(\kvt) 
\EEQ
where the $\om'_r$ are the poles of the Green function $\Gv'$ in the Lehmann representation (\ref{eq:qmatrix1}) and the
$\om_r(\kvt)$ are the poles of the VCA Green function $(\Gv_0^{-1}(\kvt)-\Sigmav)^{-1}$.
The latter are the eigenvalues of the $R\times R$ matrix $\Lv(\kvt) = \Lambdav + \Qv^\dg\Vv(\kvt)\Qv$ (see Appendix~\ref{subsec:Lehmann}).
$R$ is the number of columns of the Lehmann representation matrix $\Qv$, basically the total number of iterations performed in the band Lanczos procedure.

In practice, the first sum in (\ref{eq:omega4}) is readily calculated.
The second sum demands an integration over wavevectors.
For each wavevector $\kvt$, one must calculate $\Lv(\kvt)$ and find its eigenvalues, a process of order $R^3$.
Other linear-algebraic manipulations leading to the diagonalization of $\Lv(\kvt)$ are typically less time-consuming than the diagonalization itself.
The computation time therefore goes like $N_k R^3$, where $N_k$ is the number of points in a mesh covering the reduced Brillouin zone (in fact half of the reduced Brillouin zone, since inversion symmetry is assumed).

\subsubsection{Method II : Numerical frequency integration}\label{subsec:NI}

An alternate method of computing the sums in (\ref{eq:sef4}) is to perform them in the reverse order, i.e., to first compute the wavevector sum for a fixed frequency $\om$, and then integrating over frequencies numerically.
The method used to sum over wavevectors is exactly the same as in Method I above : a wavevector mesh is set up in the reduced Brillouin zone.
This mesh is either a fixed, regular grid, or an adaptive mesh that is refined recursively as needed by comparing a two- and three-points Gauss-Legendre evaluations within each cell (more accurately, the number of function evaluations in each cell is $2^d$ and $3^d$, $d$ being the dimension of space).

In the limit of zero temperature, the second term of the Potthoff functional (\ref{eq:sef4}) may be written as
\BEQ\label{eq:freqInt}
I = \int_C{\dr\om\over 2\pi i} \frac{L}{N}\sum_\kvt\ln\det\Big[1-\Vv(\kvt)\Gv'(\om)\Big] 
\EEQ
where the frequency integral $I$ is carried along a closed, counterclockwise contour $C$ that encloses the negative real axis, following the usual prescription with Green functions.
We show in Appendix~\ref{section:frequencyIntegral} that this integral reduces to
\BEQ
I = \int_0^\infty {\dr x\over\pi} \frac{L}{N}\sum_\kvt  \ln\Big|\det(1-\Vv(\kvt)\Gv'(ix))\Big| - L(\mu-\mu')
\EEQ
We refer to Appendix \ref{section:frequencyIntegral} for details, and for a discussion of the merits of this approach compared to the exact method above.
This is the method that we generally follow.

\subsection{Example : Antiferromagnetism}

Let us start our examples with N\'eel antiferromagnetism.
The corresponding Weiss field is defined in (\ref{eq:weissAF}).
Fig.~\ref{fig:AF1} shows the Potthoff functional as a function of N\'eel Weiss field $M$ for various values of $U$, at half-filling, calculated on a $2\times2$ cluster.
We note three solutions per curve: two equivalent minima located symmetrically about $M=0$, and a maximum at $M=0$ corresponding to the normal state solution.
The normal and AF solutions both correspond to half-filling, and the AF solution has a lower energy density $\Ec = \Om + \mu n$.
We therefore conclude, on this basis, that the system has AF long-range order.
Note that, as $U$ is increased, the profile of the curve is shallower and the minimum closer to zero.
Indeed, for large $U$, the half-filled Hubbard model is well approximated by the Heisenberg model with exchange $J=4t^2/U$, and the curve should (and will) scale towards a fixed shape when $\Omega/J$ is plotted against $M/J$ (both dimensionless quantities).
Fig.~\ref{fig:AF3} shows how the optimal Weiss field and the N\'eel order parameter vary as a function of $U$.
The Weiss field vanishes both as $U\to0$, where the order disappears, and as $U\to\infty$.
In both limits the energy difference between normal and broken symmetry state (or `condensation energy') goes to zero (Fig.~\ref{fig:AF3}), and so should the critical (N\'eel) temperature.
The order parameter $\L\Oc_M\R$ increases monotonically with $U$ and saturates.

\InsertFigure{Potthoff functional as a function of N\'eel Weiss field $M$ for various values of $U$, at half-filling, calculated on a $2\times2$ cluster. The positions of the minima are indicated.\label{fig:AF1}}{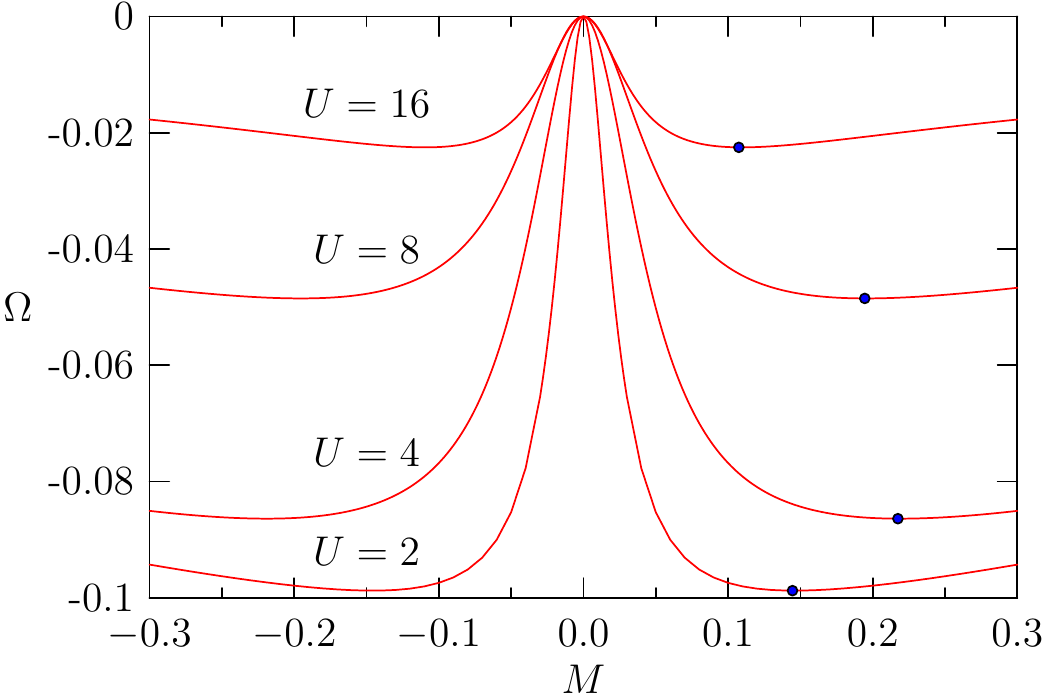}{0.9\hsize}

\InsertFigure{Optimal N\'eel Weiss field $M$ and corresponding order parameter, as a function of $U$, at half-filling, calculated on a $2\times2$ cluster.
Also shown is the ordering energy, i.e., the difference between the energy density of the normal state and that of the N\'eel state (in fact the difference between the grand potentials of the two solutions, since they both sit at half-filling).\label{fig:AF3}}{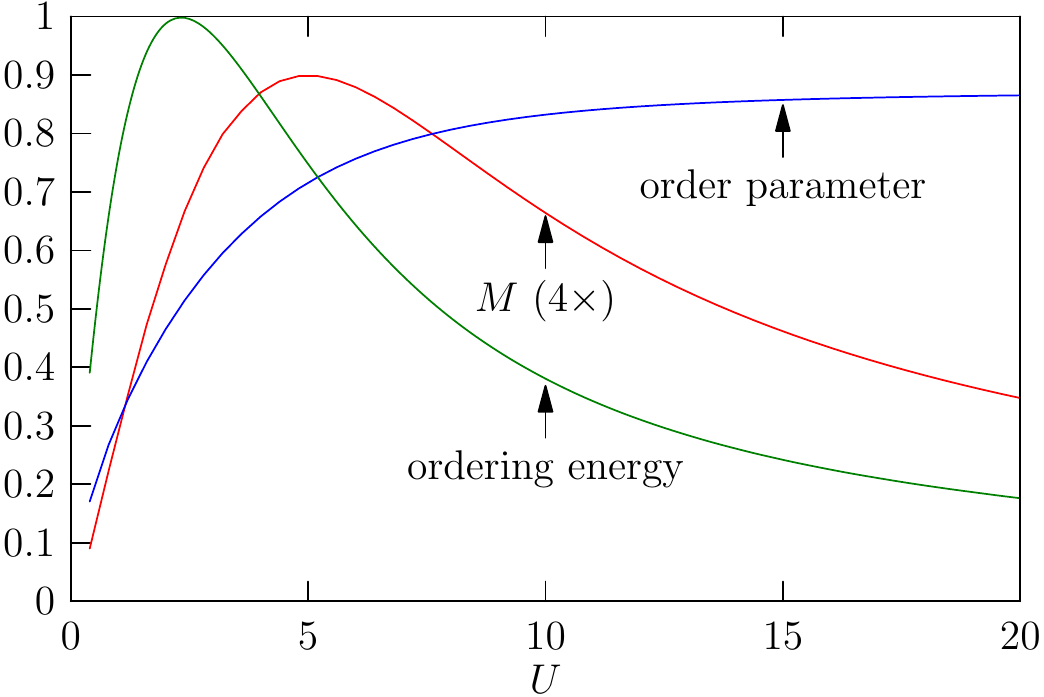}{0.9\hsize}

\InsertFigure{Potthoff functional as a function of N\'eel Weiss field $M$ for various cluster sizes, at half-filling and $U=8$.
The clusters used (from top to bottom) are: $2\times2$, $2\times3$, $2\times4$, B10 -- see Fig.~\ref{fig:B10} --, and $3\times4$. The positions of the minima are indicated.\label{fig:AF2}}{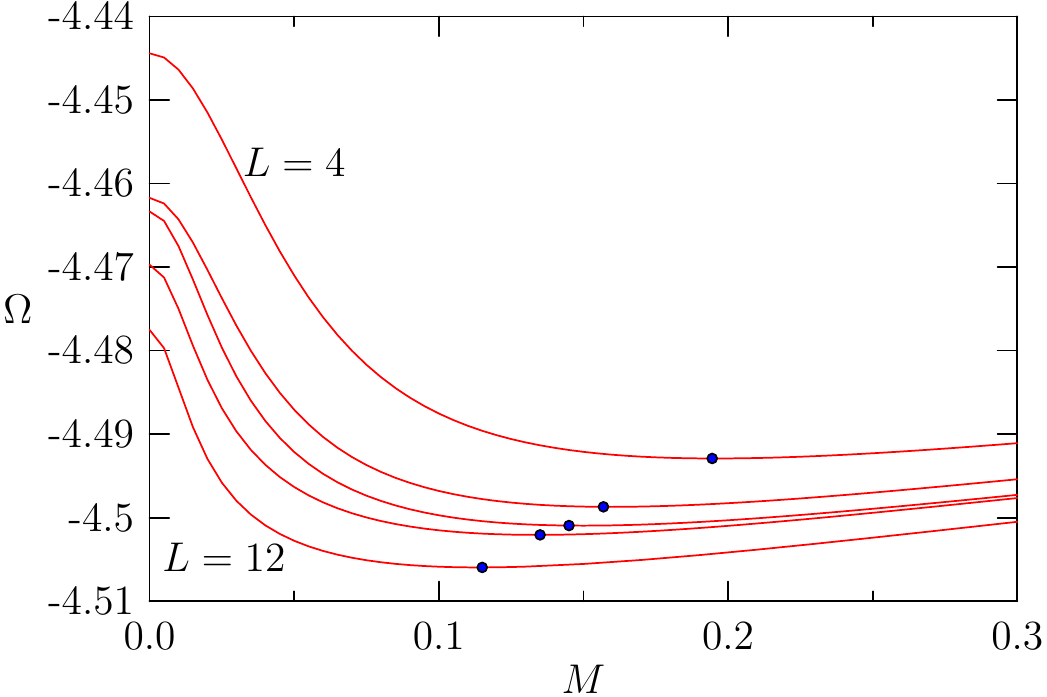}{0.9\hsize}

\begin{figure}
\begin{center}
\includegraphics[width=0.9\hsize]{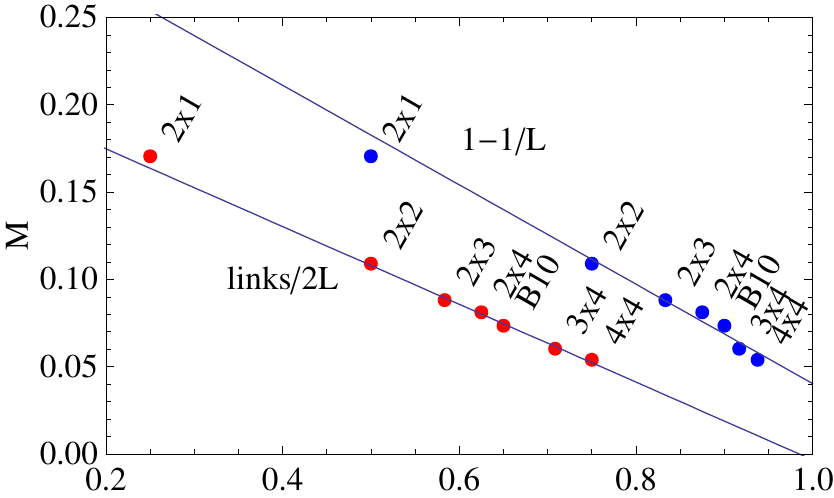}
\includegraphics[width=0.9\hsize]{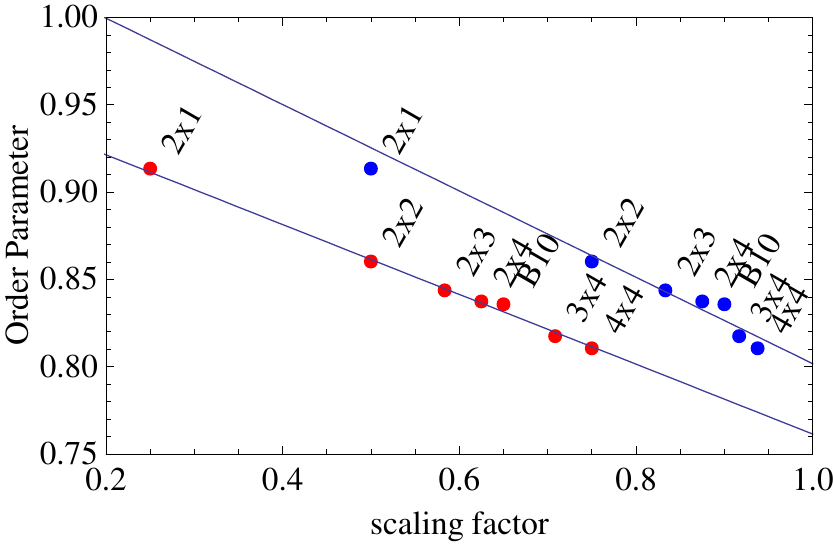}
\caption{Top: Optimal N\'eel Weiss field for the half-filled Hubbard model at $U=16$, as a function of scaling parameter.
Blue points : the scaling parameter is $1-1/L$, and the scaling is poor.
Red points : the scaling parameters is the number of cluster links divided by $2L$ -- this takes open boundary conditions into account. We see how the Weiss field goes to zero in the thermodynamic limit. Bottom: Same, for the N\'eel order parameter, which tends to a finite value in the thermodynamic limit. Against the second scaling parameter works better.
\label{fig:scaling}}
\end{center}
\end{figure}

Fig.~\ref{fig:AF2} shows the Potthoff functional as a function of N\'eel Weiss field $M$ for various cluster sizes, at half-filling and $U=8$.
There is a clear and monotonous size dependence of the position of the minimum.
In particular, the optimal Weiss field decreases as cluster size increases.
This should not worry us, quite on the contrary.
The Weiss field is needed only because spontaneously broken symmetries cannot arise on a finite cluster.
The bigger the cluster, the easier it is to break the symmetry and the optimal Weiss field should tends towards zero as the cluster size goes to infinity.
Finite-size scaling is generally very difficult, because cluster sizes are small and clusters vary in shape as well as size.
Moreover, open boundary conditions are used rather than periodic ones, which adds edge effects to size effects.
One needs to define a scaling parameter $q$, ranging between 0 and 1, that somehow defines the ``quality'' of the cluster ($q=1$ being the thermodynamic limit).
Fig.~\ref{fig:scaling} shows the optimal N\'eel Weiss field as a function of two possibilities for the scaling factor $q$, for the half-filled Hubbard model at $U=16$.
The first possibility (blue dots) is $q=1-1/L$, which does not take into account the shape of the cluster.
The second possibility (red dots) corresponds to $q$ defined as the number of links on the cluster, divided by twice the number of sites.
This also goes to 1 in the thermodynamic limit (for the square lattice), but this time takes into account the boundary of the cluster.
Indeed, $1-q$ corresponds to the fraction of links of the lattice that are ``inter-cluster'' and thus treated ``perturbatively'' in the CPT sense.
In that case, the scaling is good, as the optimal Weiss fields extrapolates very close to zero in the $q\to 1$ limit.
At the same time, the AF order parameter also decreases, bu extrapolates to a finite value, as shown on the same figure

\subsection{Superconductivity}

Superconductivity requires the use of pairing fields as Weiss fields, i.e., of operators creating Cooper pairs at specific locations.
Generally, pairing fields have the form
\BEQ
\Oc_{\rm sc} = \sum_{\rv\rv'} \Delta_{\rv\rv'} c_{\rv\up}c_{\rv'\dn} + {\rm H.c}
\EEQ
Different types of superconductivity correspond to different pairing functions $\Delta_{\rv\rv'}$.
For instance, ordinary (local) $s$-wave pairing (\textit{\`a la} BCS) corresponds to $\Delta_{\rv\rv'} = \delta_{\rv\rv'}$.
On a square lattice, what is usually known as $d_{x^2-y^2}$ pairing corresponds to 
\BEQ\label{eq:dx2y2}
\Delta_{\rv\rv'} = \l\{\ALIGNED{&1~&\text{if}~ \rv-\rv' = \pm\ev_x \\ -&1 &\text{if}~ \rv-\rv' = \pm\ev_y}\r.
\EEQ
whereas $d_{xy}$ pairing corresponds to 
\BEQ
\Delta_{\rv\rv'} = \l\{\ALIGNED{&1~&\text{if}~ \rv-\rv' = \pm(\ev_x+\ev_y) \\ -&1 &\text{if}~ \rv-\rv' = \pm(\ev_x-\ev_y)}\r.
\EEQ
The above two pairing are spin singlets.

Pairing fields, once introduced in the cluster Hamiltonian $H'$ as Weiss fields, do not conserve particle number (but conserve spin).
This increases the computational burden, since now the Hilbert space must be increased to include all sectors of a given total spin.
In practice, one uses the Nambu formalism, which in this case amounts to a particle-hole transformation for spin-down operators.
Indeed, if we introduce the operators
\BEQ
c_\rv = c_{\rv\up} \Text{and} d_\rv = c_{\rv\dn}^\dg
\EEQ
then the pairing fields look like simple hopping terms between $c$ and $d$ electrons, and the whole cluster Hamiltonian can be kept in the standard form (\ref{eq:Hubbard}), albeit with hybridization between $c$ and $d$ orbitals.

\InsertFigure{Profile of the Potthoff functional as a function of Weiss field for various superconducting pairing fields.
The extended $s$-wave is defined as the same as (\ref{eq:dx2y2}), but without the sign change between $x$ and $y$ directions.\label{fig:SC}}{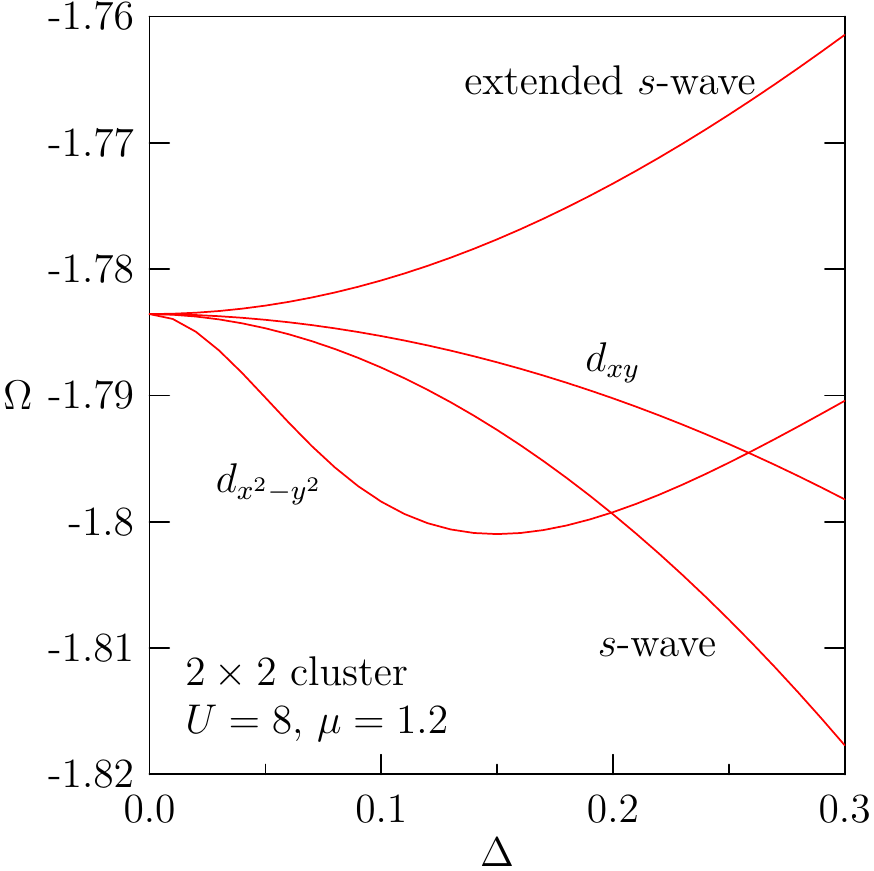}{0.75\hsize}

Fig.~\ref{fig:SC} illustrates the dependence of the Potthoff functional on various superconducting pairing fields (generically denoted $\Delta$).
In that case, only $d_{x^2-y^2}$ pairing leads to a nontrivial solution.
Others are piece-wise monotonously increasing or decreasing function, with a single zero-derivative point at $\Delta=0$.

\subsection{Thermodynamic consistency}

\InsertFigure{Comparisons of the estimates of the electron density $n$ as a function of chemical potential $\mu$, with different methods of calculation, for the normal solution, at $U=8$, on a $2\times2$ cluster. The subscript `cons.' means that the corresponding quantities were computing in a thermodynamically consistent way, by using $\mu'$ as a variational parameter.\label{fig:coherence}}{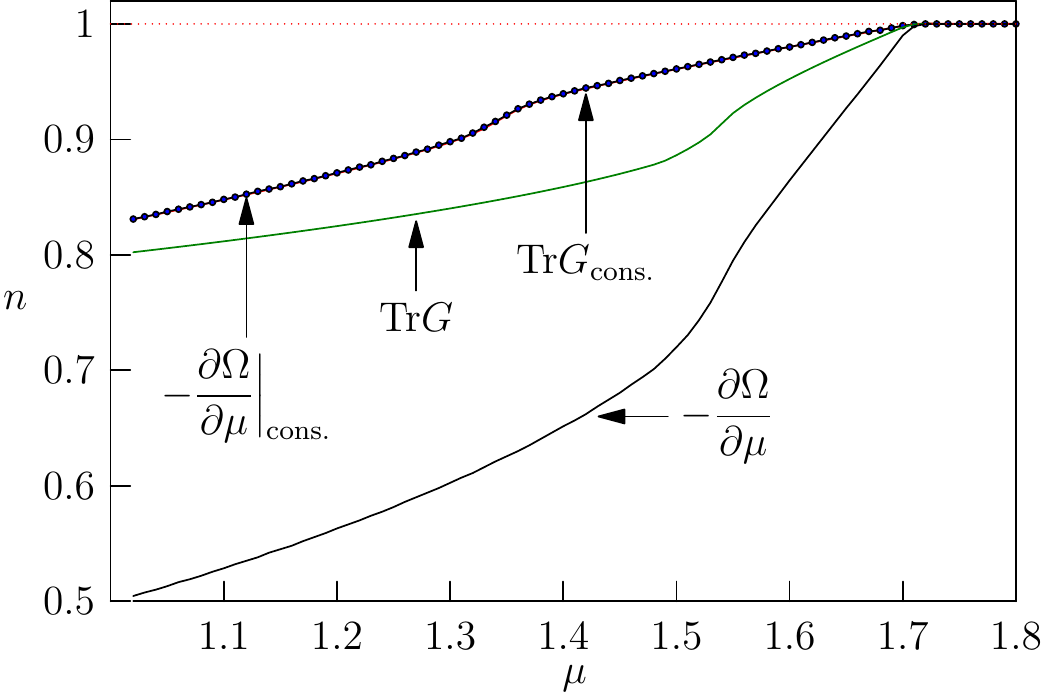}{0.8\hsize}

One of the main difficulties associated with VCA (or CPT) is the limited control over electron density.
In the absence of pairing fields, electron number is conserved and clusters have a well-defined number of electrons.
This makes a continuously varying electron density a bit hard to represent.
Of course, one may simply vary the chemical potential $\mu$ and look at the corresponding variation of the electron density, given by the trace of the Green function (schematically, $\Tr\Gv$, see Appendix~\ref{section:averages}).
This provides a continuously varying estimate of the density as a function of $\mu$.
An alternate way of estimating the density is to use the relation
\BEQ
n = -\pdel\Omega\mu
\EEQ
where the grand potential $\Om$ is approximated by the Potthoff functional at the solution found, and $\mu$ is varied as an external parameter.
The problem is that the two estimates do not coincide (see Fig.~\ref{fig:coherence}).
In other words, the approach is not thermodynamically consistent.
The recipe to make it consistent is simple : the chemical potential $\mu'$ of the cluster should not be assumed to be the same as that of the lattice system ($\mu$), but should be treated as a variational parameter.
If this is done, then the two methods for calculating $n$ given precisely the same result (see Fig.~\ref{fig:coherence}), and this can easily be proven in general.
Results on a Hubbard model for the cuprates with thermodynamic consistency are shown on Fig.~\ref{fig:SC3x4}; see also Ref.~\onlinecite{Aichhorn:2006rt}.

\InsertFigure{Order parameters for $d_{x^2-y^2}$ pairing and N\'eel antiferromagnetism for a model of the high-$T_c$ cuprates with $U=8$, diagonal hopping $t_1=-0.3$ and third neighbor hopping $t_2=0.2$. Calculations are performed on a $3\times4$ cluster.
Three solutions are displayed: (1) a pure $d_{x^2-y^2}$, obtained with two variational parameters ($\mu'$ and $\Delta_{x^2-y^2}$);
(2) a pure N\'eel solution obtained by varying $\mu'$ and the N\'eel Weiss field $M$; a homogeneous coexistence solution obtained by varying $\mu'$, $M$ and $\Delta_{x^2-y^2}$ (From M. Guillot, MSc thesis, Universit\'e de Sherbrooke).
\label{fig:SC3x4}}{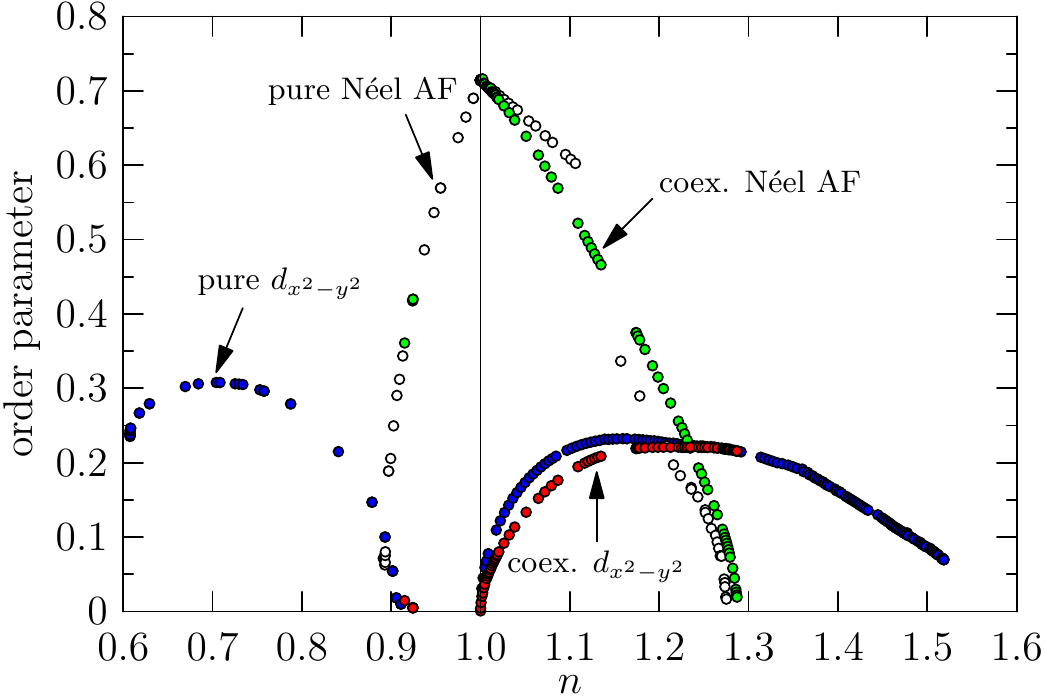}{0.99\hsize}

\subsection{Searching for stationary points}

Let $x_i$ be the $n$ different variational parameters used in VCA.
Once the function $\Omega(\xiv)$ may be efficiently calculated, it remains to find a stationary point of that function.
This point is not necessarily a minimum in all directions.
Indeed, experience has shown that $\om$ is a maximum as a function of the cluster chemical potential $\mu'$, while it is generally a minimum as a function of symmetry-breaking Weiss fields like $M$ or $\Delta$.

The Newton-Raphson algorithm allows one to find stationary points with a small number of function evaluations.
One starts with a trial point $\xiv_0$ and an initial step $h$.
Let $\ev_i$ denote the unit vector in the direction of axis $i$ of the variational space.
The function $\om$ is then calculated at as many points as necessary to fit a quadractic form in the neighborhood of $\xiv_0$.
This requires $(n+1)(n+2)/2$ evaluations, at points like $\xiv_0$, $\xiv_0\pm h\ev_i$, and a few of $\xiv_0 + h(\ev_i + \ev_j)$.
The stationary point $\xiv_1$ of that quadratic form is then used as a new starting point, the step $h$ is reduced to a fraction of the difference $|\xiv_1-\xiv_0|$, and the process is iterated until convergence on $|\xiv_i-\xiv_{i-1}|$ is achieved.
A variant of this method, the quasi-Newton algorithm, may also be used, in which the full Hessian matrix of second derivatives is not calculated. It requires in general more iterations, but fewer function evaluations at each step.

The advantage of the Newton-Raphson method lies in its economy of function evaluations, which are very expensive here: each requires the solution of the cluster Hamiltonian.
Its disadvantage is a lack of robustness.
One has to be relatively close to the solution in order to converge towards it.
But one typically runs parametric studies in which an external (i.e. non variational) parameter of the model is varied, such as the chemical potential $\mu$ or the interaction strength $U$.
In this context, the solution associated with the current value of the external parameter may be used as the starting point for the next value, and in this fashion, by proximity, one may conduct rather robust calculations.

A more robust method, albeit more time consuming, is the conjugate-gradient algorithm, which we will not explain here as it is amply documented and fairly common.
However, this algorithm finds minima (or maxima), not saddle points in general.
We must therefore take the extrinsic step of identifying parameters (like $\mu'$ above) that are expected to drive maxima of $\om$, and a complementary set of parameters (like $M$ and $\Delta$ above) that drive minima of $\om$.
One then, iteratively, finds maxima and minima with the two sets of parameters in succession, and stops when convergence on $|\xiv_i-\xiv_{i-1}|$ has been achieved.
This method is suitable to find a first solution when the Newton-Raphson method fails to deliver one.
It may however converge to minima that are in fact singularities of $\om$, i.e., points where the derivatives are not defined.
Such points may occur as the result of energy-level crossings in clusters and are an artifact of the finite-cluster size.

\section{The Cellular Dynamical Mean Field Theory}

The Cellular dynamical mean-field theory (CDMFT) -- also called Cluster dynamical mean-field theory -- is a cluster extension of Dynamical mean-field theory (DMFT).
Since there is no real pedagogical gain in describing first DMFT, we will proceed directly to CDMFT, in the context of a an exact diagonalization solver.

\InsertFigure{Examples of clusters with baths. Bath sites are square, cluster sites blue circles.
Bath parameters for the normal solution are indicated on the two top clusters, while conventional labels of orbitals are indicated for the (4+8)-site cluster.\label{fig:bath}}{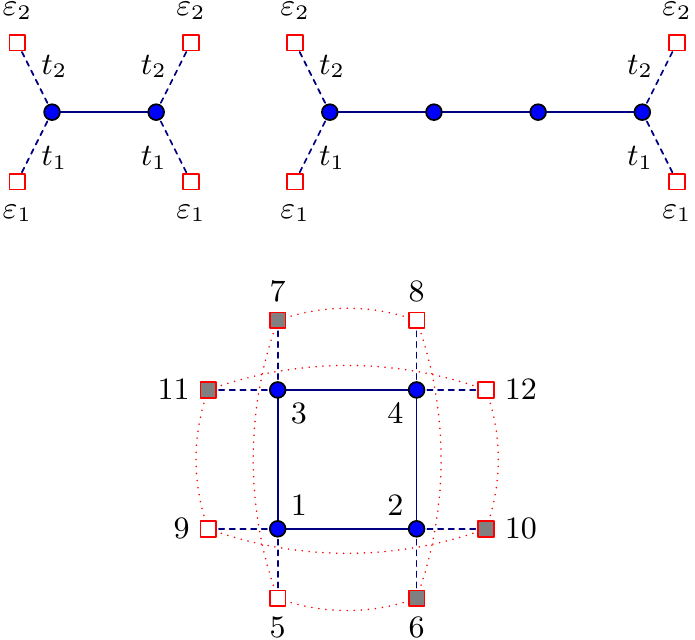}{0.9\hsize}

The basic idea behind CDMFT is to model the effect on the cluster of the remaining degrees of freedom of the lattice by a bath of uncorrelated orbitals that exchange electrons with the cluster, and whose parameters are set in a self-consistent way.
Explicitly, the cluster Hamiltonian $H'$ takes the form
\ALIGN{\label{eq:cdmft1}
H' =&-\sum_{\mu,\nu}t_{\mu\nu}c_\mu^\dg c_\nu + U\sum_{\Rv}n_{\Rv\up}n_{\Rv\dn}\notag \\
&+\sum_{\mu,\a}\t_{\mu\a}(c_\mu^\dg a_\a + \mathrm{H.c.}) + \sum_\a \eps_\a a_\a^\dg a_\a
}
where $a_\a$ annihilates an electron on a bath orbital labelled $\a$.
The label $\a$ includes both an `bath site' index and a spin index for that `site'.
The bath is characterized by the energy of each orbital ($\eps_\a$) and by the bath-cluster hybridization matrix
$\t_{\mu\a}$ (the index $\mu$ includes cluster site, spin and band indices).
This representation of the environment through an Anderson impurity model was introduced in Ref.~\onlinecite{Caffarel:1994} in the context of DMFT (i.e., a single-site cluster).
Note that `bath site' is a misnomer, as bath orbitals have no position assigned to them.

The effect of the bath on the electron Green function is encapsulated in the so-called hybridization function
\BEQ\label{eq:hybridization}
\Gamma_{\mu\nu}(\om) = \sum_\a \frac{\t_{\mu\a}\t^*_{\nu\a}}{\om-\eps_\a}
\EEQ
which enters the electron Green function as
\BEQ
\Gv'{}^{-1} = \om - \tv - \Gammav(\om) - \Sigmav(\om)
\EEQ
This is shown in Appendix~\ref{section:bath} in the non-interacting case ($\Sigmav=0$).
By definition, the only effect of adding the electron-electron interaction is to add the self-energy $\Sigmav$, as above.

\subsection{Bath degrees of freedom and SFA}

The CDMFT Hamiltonian (\ref{eq:cdmft1}) defines a valid reference system for Potthoff's self-energy functional approach, since it shares the same interaction part as the lattice Hamiltonian $H$ and since each cluster of the superlattice has its own identical, independent copy.
From the SFA point of view, the bath parameters $\{\eps_\a,\t_{\mu\a}\}$ can in principle be chosen in such a way as to make the Potthoff functional stationary.
A subtlety arises: the bath system must be considered part of the original Hamiltonian $H$, albeit without hybridization to the cluster sites, in order for both Hamiltonians to describe the same degrees of freedom; but within $H$ we are free to give the bath trivial parameters ($\eps_\a=0$).
Performing VCA-like calculations with bath degrees of freedom is illustrated in Ref.~\onlinecite{Balzer:2008}, and on Fig.~(\ref{fig:1D-2-4b}) below.

When evaluating the Potthoff functional in the presence of a bath, one must add a contribution from the bath to $\Tr\ln(-\Gv')$, which takes the form
\BEQ
\Omega_{\rm bath} = \sum_{\eps_\a < 0} \eps_\a
\EEQ
and which comes from the zeros of the cluster Green function induced by the poles of the hybridization function.
Note that the zeros coming from the self-energy cancel out in Eq.~(\ref{eq:omega4}) between the contribution of $\Tr\ln(-\Gv')$ and that of
$\Tr\ln(-\Gv)$, but not those coming from $\Gammav(\om)$, as they only occur in $\Gv'$.

\InsertFigure{Electron density as a function of chemical potential, for the one-dimensional Hubbard model, at $U=3$.
The black curve is the exact result from the Lieb-Wu solution using the Bethe Ansatz.
The red curve is the SFA calculation on the 2-site cluster with 4 bath sites shown on Fig.~\ref{fig:bath}.
The other curves are obtained with the CDMFT algorithm (parameters explained in the text).\label{fig:1D-2-4b}}{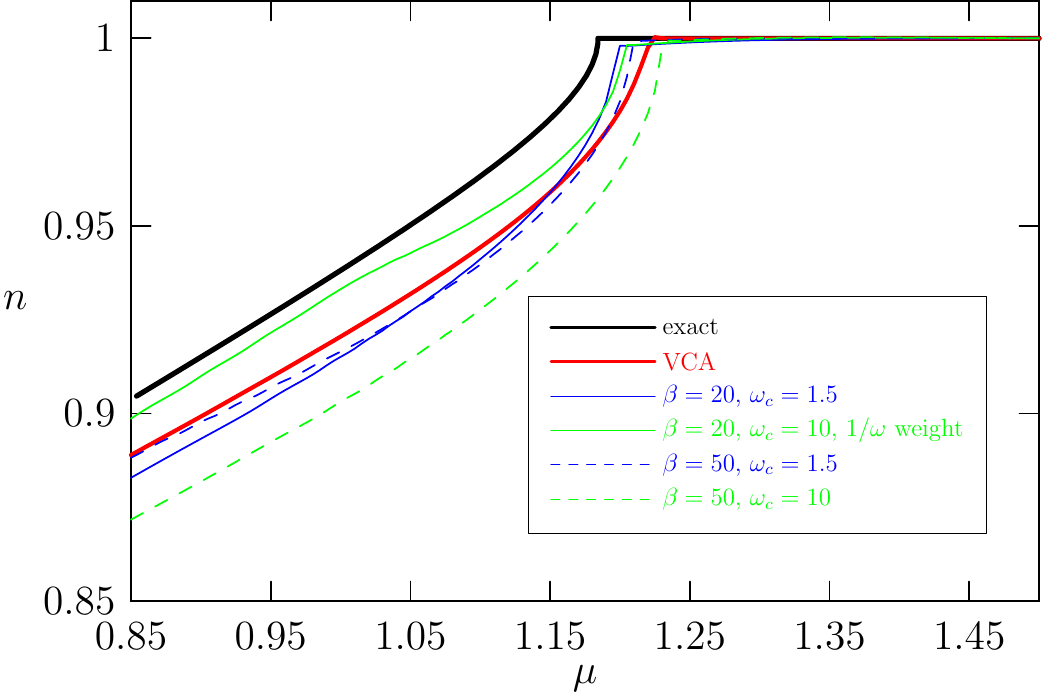}{0.99\hsize} 

\InsertFigure{Two of the four bath parameters used in the calculations of Fig.~\ref{fig:1D-2-4b}, as a function of $\mu$. The legend is the same as for Fig.~\ref{fig:1D-2-4b}.\label{fig:1D-2-4b-param}}{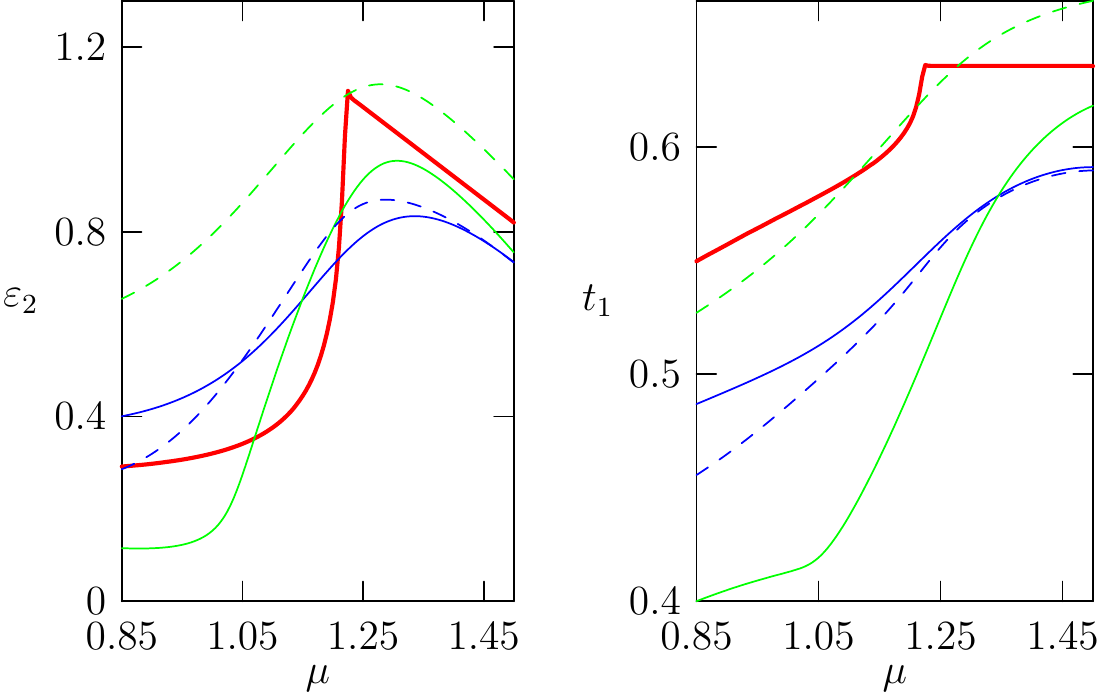}{0.99\hsize} 

\subsection{The CDMFT self-consistent procedure}

\InsertFigure{The CDMFT algorithm with an exact diagonalization solver.\label{fig:cdmftAlgo}}{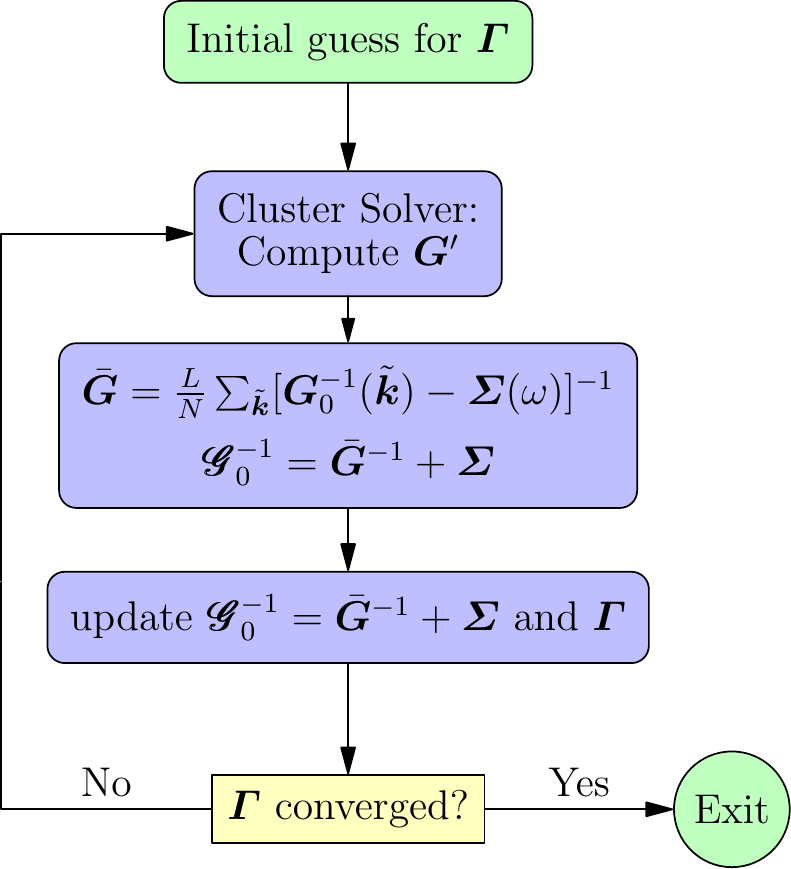}{0.8\hsize}

However, in practice, CDMFT does not proceed in this way, i.e., it does not look for a strict solution of the Euler equation
(\ref{eq:EulerVCA}).
It tries instead to set each of the terms between brackets to zero separately.
Since the Euler equation (\ref{eq:EulerVCA}) can be seen as a scalar product, CDMFT requires that the modulus of one of the vectors vanish to make the scalar product vanish.
From a heuristic point of view, it is as if each component of the Green function in the cluster were equal to the corresponding component deduced from the lattice Green function.
Clearly, the left-hand side of Eq.~(\ref{eq:EulerVCA}) cannot vanish separately for each frequency, since the number of degrees of freedom in the bath is insufficient.
Instead, one adopts the following self-consistent scheme (see Fig.~\ref{fig:cdmftAlgo}):
\BE
\item Start with a guess value of the bath parameters $(\t_{\mu\a},\eps_\a)$, that define the hybridization function (\ref{eq:hybridization}).
\item Calculate the cluster Green function $\Gv(\om)$ with the Exact diagonalization solver.
\item Calculate the superlattice-averaged Green function
\BEQ
\bar\Gv(\om)=  \frac LN\sum_{\kvt}\frac{1}{\Gv_{0}^{-1}(\kvt)-\Sigmav(\om)}
\EEQ
and the combination
\BEQ
\Gcv_{0}^{-1}(\om)=\bar\Gv^{-1}+\Sigmav(\om)
\EEQ
\item Minimize the following distance function:
\BEQ\label{dist_func}
d=\sum_{\om,\nu,\nu'}\l\vert\l(\om + \mu-\tv'-\Gammav(\om)-\Gcv_0^{-1}\r)_{\nu\nu'}\r\vert^{2}
\EEQ
over the set of bath parameters (changing the bath parameters at this step does not require a new solution of the Hamiltonian $H'$, but merely a recalculation of the hybridization function $\Gammav$).
\item Go back to step (2) with the new bath parameters obtained from this minimization, until they are converged.
\EE

In practice, the distance function (\ref{dist_func}) can take various forms,
for instance by adding a frequency-dependent weight in order to emphasize low-frequency properties\cite{Kancharla:2008,Bolech:2003ye, Stanescu:2006fk} or by using a sharp frequency cutoff.\cite{Kyung:2005} These weighting factors can be considered as rough approximations for the missing factor $\delta\Sigma_{\nu\mu}'(\om)/\delta\tv'$ in the Euler equation (\ref{eq:EulerVCA}).
The frequencies are summed over on a discrete,
regular grid along the imaginary axis, defined by some fictitious inverse temperature $\beta$, typically of the order of 20 or 40 (in units of $t^{-1}$).
Even when the total number of cluster plus bath sites in CDMFT equals the number of sites in a VCA calculation, CDMFT is much faster than the VCA since the minimization of a grand potential functional requires many exact diagonalizations of the cluster Hamiltonian $H'$.

\subsection{Examples}

Let us start with a one-dimensional example.
Fig.~(\ref{fig:1D-2-4b}) illustrates the variability of CDMFT results related to the choice of the distance function.
The cluster used has two sites and four bath sites (see Fig.~\ref{fig:bath}), and the various curves represent the electron density $n$ as a function of chemical potential $\mu$ for the one-dimensional Hubbard model at $U=3$.
The exact results, from the Lieb-Wu solution, is shown in black, as well as the SFA result coming from an exact solution of the Euler equation (\ref{eq:EulerVCA}) for that system.
The four CDMFT results shown differ by the value of $\beta$ and that of the sharp frequency cutoff $\om_c$.
In addition, one of the curves was obtained by weighing the different frequencies by a factor $1/|\om|$.
An important characteristic of the exact result is that infinite compressibility $\d n/\d\mu$ at the point where the gap ends, i.e., when the density curve hits $n=1$, at a value $\mu_c(U)$ of the chemical potential. The SFA and CDMFT do quite well in accounting for the infinite compressibility, contrary to other approaches (e.g. one-site DMFT).
On this small two-site system, they do not find the correct $\mu_c$, but increasing the bath size would improve on this.
By playing with the distance function, on may bring the curves closer to or further from the exact result, but there is no guarantee that the most successful distance function in this case will be as profitable when the exact solution is unknown!
In principle, the SFA curve (red) is the one that best represents what can be achieved with this system, and the various CDMFT curves are to be judged against not the exact result, but against the SFA curve.

\InsertFigure{N\'eel (AF) and $d$-wave (dSC) order parameters obtained from CDMFT applied to the (4+8)-cluster of Fig.~\ref{fig:bath},
for the two-dimensional Hubbard model with $U=8$ and diagonal hopping $t'=-0.3$.
The data is shown as a function of the calculated lattice density $n$.
The order parameters are calculated using the same operators as in the corresponding VCA calculation illustrated on Fig.~\ref{fig:SC}, even though these operators played no role in the solution: they are merely used as a probe.
In this calculation, we set $\beta=20$ and a sharp cutoff $\om_c=3$ was used.
The dSC and AF solutions were both allowed simultaneously (9 bath parameters) and there are regions of coexistence of the two orders.
\label{fig:2x2-8b}}{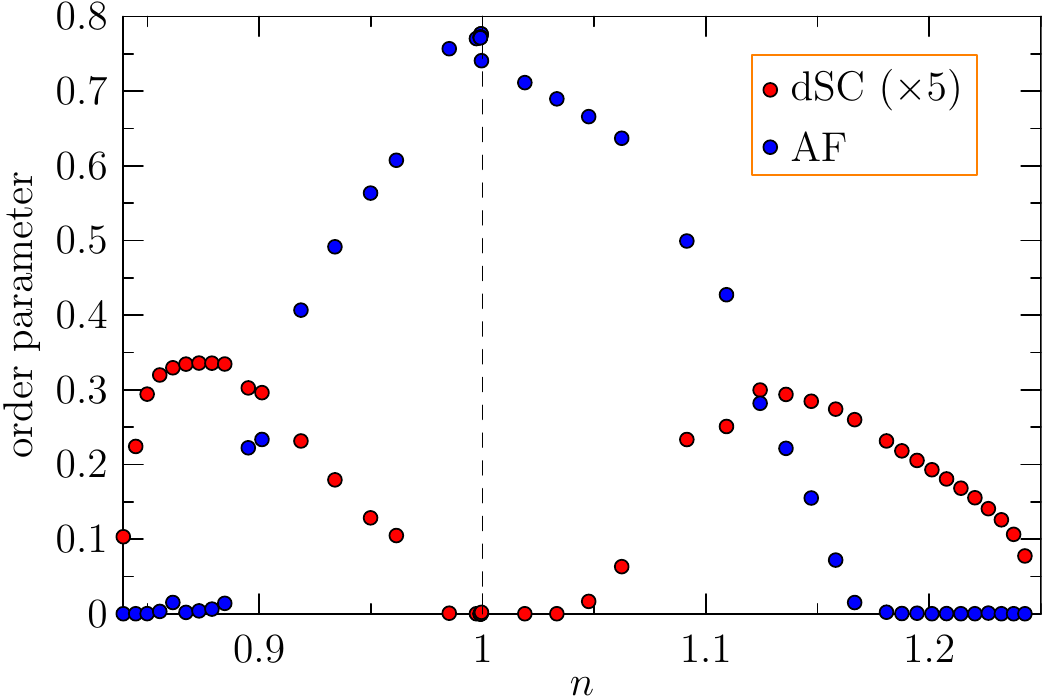}{0.9\hsize}

Next, consider the two-dimensional cluster illustrated in the lower part of Fig.~\ref{fig:bath}.
This 4-site, 8-bath site cluster is the main cluster used in CDMFT simulations of high-$T_c$ cuprates using the two-dimensional Hubbard model.
It is useful in that case to view the orbitals numbered 5 to 8 as a first bath set, and the orbitals numbered 9 to 12 as a second bath.
Each site of the cluster is connected to one orbital of each set.
In studying the normal state, and taking into account the symmetries of the cluster, we would need 4 bath parameters: one bath-cluster hopping and one bath energy for each set.
In order to treat a possible antiferromagnetic phase, one must modify the bath energies and hopping in a spin-dependent way.
The grey and white squares on the figure then distinguish orbitals of a given bath according to their shift in site energy (of opposite signs for opposite spins).
The corresponding bath-cluster hybridization may also be different, which makes a total of 8 parameters.
Finally, in order to study $d$-wave superconductivity, we introduce pairing within each bath (red dotted lines on the figure), vertical and horizontal pairing being of opposite signs.
This introduces an additional parameter, for a total of 9.
At this point, an important remark is in order : Formula (\ref{eq:hybridization}) for the hybridization function only applies if the bath orbitals are not hybridized between themselves.
The $d$-wave pairing just described certainly breaks that condition.
This is not a problem, however, if we perform a change of variables within bath degrees of freedom (a Bogoliubov transformation) prior to solving the problem numerically, such as to make the bath Hamiltonian diagonal.
Then the poles of the hybridization function no longer correspond to the bath energies as defined originally in the model, but rather to the eigenvalues of the bath Hamiltonian.

Results of a CDMFT calculation on this system are shown in Fig.~\ref{fig:2x2-8b}.
Comparing with the VCA result of Fig.~\ref{fig:SC3x4}, we notice first the similarities: the existence of a dSC phase away from half-filling for both electron and hole doping and the possibility of homogeneous coexistence between antiferromagnetism and d-wave superconductivity.
But differences are obvious : the VCA diagram is more asymmetric than the CDMFT one in terms of electron vs hole doping.
Both calculations agree on the critical doping for antiferromagnetism on the hole-doped side ($\sim$ 10\%), but not on the electron-doped side.
The VCA result does not show homogeneous coexistence between AF and dSC on the hole-doped side -- although it appears on smaller clusters.
At this point it is not clear whether these differences arise because of the methods themselves rather that the particular way they were applied (choice of Weiss fields, bath configuration, distance function, etc.).
In particular, the exact SFA result for the system used in CDMFT has not yet been calculated.

\appendix 
\section{Clusters and Kinematics}

In this appendix we will review the kinematics of cluster decompositions, and introduce notation used throughout this paper.
The spatial dimension $D$ of the lattice will be left general.

Cluster methods are based on a cluster decompostion of the model, i.e., on a tiling of the original lattice $\g$ with identical clusters of $L$ sites each.
Mathematically, this corresponds to introducing a superlattice $\G$, whose sites form a subset of the lattice $\g$ and will be labelled by vector base positions with tildes ($\rvt$, $\rvt'$, etc).
This superlattice is generated by $D$ basis vectors $\ev_{1,\dots,D}$ belonging to $\g$, i.e., every site $\rvt$ of the superlattice may be expressed as an integer linear combination of these basis vectors.
Associated with each site of $\G$ is a cluster of $L$ sites, whose shape is not uniquely determined by the superlattice structure.
The sites of the clusters will be labelled by their vector position (capitals): $\Rv$, $\Rv'$, etc.
Each site $\rv$ of the original lattice $\g$ can be expressed in a unique way as a combination of a superlattice vector $\rvt$ and of a site $\Rv$ within the cluster: $\rv = \rvt + \Rv$.
We have the following equivalence between summations:
\BEQ
\sum_{\rv\in\g}\cdots  =  \sum_{\rvt\in\G}\sum_\Rv \cdots ~.
\EEQ
The number of sites in the cluster is simply the ratio of the unit cell volumes of the two lattices.
In $D=3$, this is 
\BEQ
L = \frac{V_\G}{V_\g} = \l| (\ev_1\pvec\ev_2)\cdot\ev_3 \r| 
\EEQ
(the above formulae can be adapted to $D=2$ by setting $\ev_3=(0,0,1)$).

The Brillouin zone of the original lattice, denoted BZ$_\g$, contains $L$ points belonging to the reciprocal superlattice $\G^*$.
The Brillouin zone of the superlattice, BZ$_\G$, has a volume $L$ times smaller than that of the original Brillouin zone.
Any wavevector $\kv$ of the original Brillouin zone can be uniquely expressed as 
\BEQ
\kv = \Kv + \kvt~,
\EEQ 
where $\Kv$ belongs both to the reciprocal superlattice \textit{and} to BZ$_\g$, and $\kvt$ belongs to BZ$_\G$ (see Fig.~
\ref{fig:superlattice}).
Thus, we have the equivalent summations \BEQ
\sum_\kv\cdots = \sum_\kvt\sum_\Kv \cdots ~.
\EEQ

The passage between momentum space and real space, by discrete Fourier transforms, can be done either directly ($\rv\leftrightarrow\kv$), or independently for cluster and superlattice sites ($\rvt\leftrightarrow\kvt$ and $\Rv\leftrightarrow\Qv$).
This can be encoded into unitary matrices $\Uv^\g$, $\Uv^\Gamma$ and $\Uv^c$ defined as follows:
\BEQ
U^\g_{\kv,\rv} = \frac1{\sqrt{N}}\er^{-i\kv\cdot\rv} ~,~
U^\Gamma_{\kvt\rvt} = \sqrt{\frac LN}\er^{-i\kvt\cdot\rvt}~,~
U^c_{\Kv,\Rv} = \frac1{\sqrt{L}}\er^{-i\Kv\cdot\Rv}
\EEQ
The discrete Fourier transforms on a generic one-index quantity $f$ are then
\BEQ
f(\kv) = \sum_\rv  U^\g_{\kv,\rv}f_\rv ~,~
f(\kvt) = \sum_\rvt  U^\Gamma_{\kvt,\rvt}f_\rvt ~,~
f_\Kv = \sum_\Rv  U^c_{\Kv,\Rv}f_\Rv
\EEQ
or, in reverse,
\BEQ
f_\rv = \sum_\kv  U_{\kv,\rv}^{\g*} f(\kv) ~,~
f_\rvt = \sum_\kvt  U_{\kvt,\rvt}^{\Gamma*} f(\kvt) ~,~
f_\Rv = \sum_\Kv  U_{\Kv,\Rv}^{c*} f_\Kv
\EEQ
where $f$ stands for a generic one-index quantity.
Quasi continuous indices, like $\kv$ and $\kvt$, are most of the time indicated between parentheses.
This notation may rightfully be deemed capricious, since the labels $\rv$ and $\kv$ take the same number $N$ of values, but we adopt it nonetheless as it helps reminding us that the values of the labels are closely separated.

These discrete Fourier transforms close by virtue of the following identities \ALIGN{
\frac{1}{N}\sum_\kv \er^{i\kv\cdot\rv} &= \delta_\rv \qquad & \frac{1}{N} \sum_\rv \er^{-i\kv\cdot\rv}&= \Delta_\g(\kv) \\
\frac{L}{N}\sum_\kvt \er^{i\kvt\cdot\rvt} &= \delta_\rvt \qquad & \frac{L}{N} \sum_\rvt \er^{-i\kvt\cdot\rvt}&= \Delta_\G(\kvt) \\
\frac{1}{L}\sum_\Kv \er^{i\Kv\cdot\Rv} &= \delta_\Rv \qquad & \frac{1}{L} \sum_\Rv \er^{-i\Kv\cdot\Rv}&= \Delta_\g(\Kv)
}
where $\delta_\rv$ is the usual Kronecker delta, used for all labels (since they are all discrete):
\BEQ
\delta_\a = \l\{ \ALIGNED{&1 \Text{if} \a=0\\ &0 \Text{otherwise}}\r. \qquad \delta_{\a\b} \equiv \delta_{\a-\b} ~,
\EEQ
and the $\Delta$'s are the so-called Laue functions:
\ALIGN{
\Delta_\g(\kv) &= \sum_{\Qv \in \g^*} \delta_{\kv+\Qv} \\
\Delta_\G(\kvt) &= \sum_{\Pv \in \G^*} \delta_{\kvt+\Pv} ~.
}
Laue functions are used instead of Kronecker deltas in momentum space because of the possibility of Umklapp processes.
Note especially that even though 
\BEQ
\delta_\kv = \delta_\kvt \delta_\Kv\qquad(\kv=\kvt+\Kv),
\EEQ
the same does not hold for the Laue functions:
\BEQ
\Delta_\g(\kv) \ne \Delta_\G(\kvt) \Delta_\g(\Kv)~.
\EEQ
Instead we have the following relations:
\ALIGN{
\Delta_\G(\kvt) &= \sum_\Kv \Delta_\g(\kvt+\Kv) \\
\Delta_\g(\kv) &= \Delta_\g(\kvt+\Kv) = \delta_\kvt \Delta_\g(\Kv) 
}
which reflect the arbitrariness in the choice of Brillouin zone of the superlattice (we use the term \textit{Brillouin zone} in a rather liberal manner, as a complete and irreducible set of wavevectors, and not as the Wigner-Seitz cell of the reciprocal lattice.)

\InsertFigure{The reduced Brillouin BZ$_\G$ zone associated with the 10-site cluster of Fig.~\protect\ref{fig:B10}.
A wavevector $\kv$ has a unique decomposition $\kv=\tilde\kv+\Qv$, where $\Qv$ is one of the $L$ elements of the reciprocal superlattice that belongs to the original Brillouin zone BZ$_\g$.\label{fig:superlattice}}{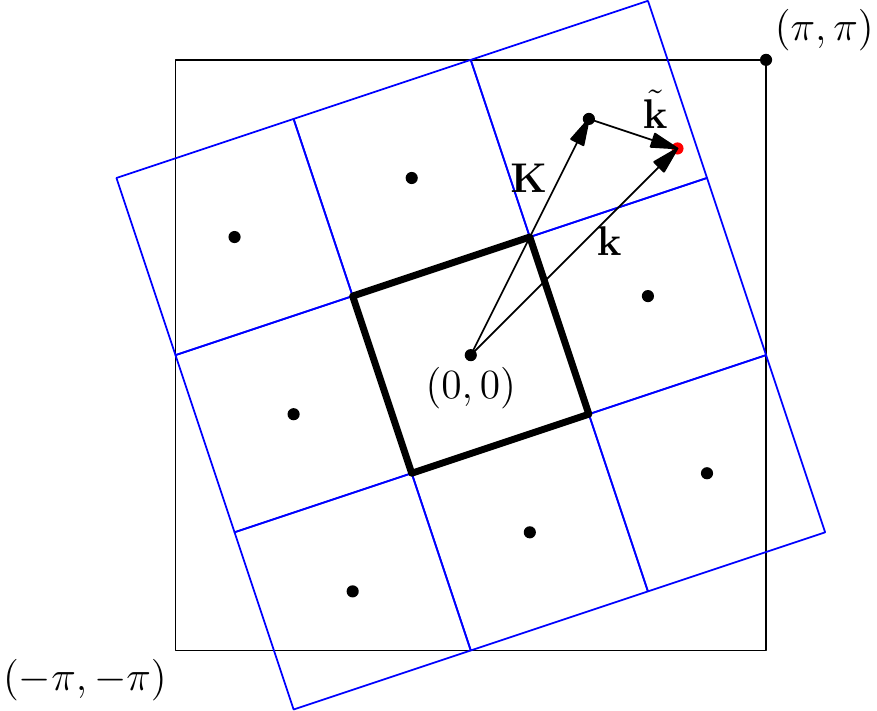}{0.8\hsize}

A one-index quantity like the destruction operator $c_\rv = c_{\rvt+\Rv}$ can be represented in a variety of ways, through partial Fourier transforms:
\ALIGN{
c_\Rv(\kvt) &= \sum_\rvt U^\Gamma_{\kvt\rvt} \, c_{\rvt+\Rv}  \\
c_{\rvt,\Kv} &= \sum_\Rv U^c_{\Kv\Rv} \, c_{\rvt+\Rv}  \\
c_\Kv(\kvt) &= \sum_{\rvt,\Rv} U^\Gamma_{\kvt\rvt}U^c_{\Kv\Rv} \, c_{\rvt+\Rv} \\
c(\kv) &= \sum_\rv U^\g_{\kv\rv} \, c_\rv
}
The last two representations are not identical, since the phases in the two cases differ by $\kvt\cdot\Rv$.
In other words, they are obtained respectively by applying the unitary matrices $\Sv\equiv\Uv^\Gamma\otimes\Uv^c$ and $\Uv^\g$ on the $\rv$ basis, and these two operations are different. 
In other words, the matrices $\Lambdav\equiv\Uv^\g\Sv^{-1}$ and $\Dv\equiv\Sv^{-1}\Uv^\g$ are not trivial:
\ALIGN{
\Lambda_{\kv\kv'} &= \delta_{\kvt\kvt'}\frac1L\sum_\Rv \er^{-i\Rv\cdot(\kvt+\Kv-\Kv')} \\
D_{\rv\rv'} &= \delta_{\Rv\Rv'}\frac LN\sum_\kvt \er^{i\kvt\cdot(\rvt-\rvt'-\Rv)}
}
and one could write
\BEQ\label{eq:lambdaMatrix}
c(\kvt+\Kv) = \sum_{\Kv'} \Lambda_{\Kv\Kv'}(\kvt) c_{\Kv'}(\kvt)
\EEQ

A two-index quantity like the hopping matrix $t_{\rv\rv'}$ may thus have a number of different representations.
Due to translation invariance on the lattice, this matrix is diagonal when expressed in momentum space: $t(\kv,\kv') = \eps(\kv) \delta_{\kv,
\kv'}$, $\eps(\kv)$ being the dispersion relation:
\BEQ
t_{\rv\rv'} = \frac1N \sum_\kv \er^{i\kv\cdot(\rv-\rv')} \eps(\kv) 
\EEQ
However, we will very often use the mixed representation 
\BEQ
t_{\Rv\Rv'}(\kvt) = \sum_\rvt \er^{i\kvt\cdot\rvt} t_{\rv\rv'} \qquad\l\{\ALIGNED{&\rv = \Rv \\ &\rv' = \rvt+\Rv'}\r.
\EEQ
For instance, if we tile the one-dimensional lattice with clusters of length $L=2$, the nearest-neighbor hopping matrix, corresponding to the dispersion relation $\eps(k) = -2t\cos(k)$, has the following mixed representation:
\BEQ
\tv(\kt) = -t\MATRIX{0 & 1 + e^{-2i\kt}\\ 1 + e^{2i\kt} & 0}
\EEQ

Finally, let us point out that the space $E$ of one-electron states is larger than the space of lattice sites $\gamma$, as it includes also spin and band degrees of freedom, which forms a set $B$ whose elements are indexed by $\s$. We could therefore write $E=\gamma\otimes B$. The transformation matrices defined above ($\Uv^\g$, $\Uv^\Gamma$ and $\Uv^c$) should, as necessary, be understood as tensor products
($\Uv^\g\otimes\id$, $\Uv^\Gamma\otimes\id$ and $\Uv^c\otimes\id$) acting trivially in $B$. This should be clear from the context.

\section{Lehmann representation of the Green function}
\label{section:Lehmann}

By inserting a completeness relations in the expression (\ref{eq:green}) for the zero-temperature Green function, one finds the Lehmann representation:
\BEQ\begin{split}
G'_{\mu\nu}(\om) = &\sum_m\L\Omega|c_\mu|m\R{1\over \om-E_m+E_0}\L m|c_\nu^\dg|\Omega\R \\ 
 + &\sum_n\L\Omega|c_\nu^\dg|n\R{1\over \om+E_n-E_0}\L n|c_\mu|\Omega\R \end{split}
\EEQ
(recall that $\mu=(\Rv,\s)$ is a compound index for cluster site and spin or band).
The two sums are over different sets of eigenstates, in the spaces with one more and one less electron, respectively.
Let us introduce the notation
\BEQ\label{eq:qmatrix0}
Q^{(e)}_{\mu m} = \L\Omega|c_\mu|m\R \Quad Q^{(h)}_{\mu n} = \L\Omega|c_\mu^\dg|n\R 
\EEQ
as well as $\om^{(e)}_m = E_m-E_0 > 0$ and $\om^{(h)}_n = -E_n+E_0 < 0$ to write
\BEQ
G'_{\mu\nu}(\om) = \sum_m{Q^{(e)}_{\mu m}Q^{(e)*}_{\nu m}\over \om-\om^{(e)}_m} + 
\sum_n{Q^{(h)}_{\mu n}Q^{(h)*}_{\nu n}\over \om-\om^{(h)}_n} 
\EEQ
The $Q^{(e)}_{\mu m}$ form a $2L\times N^{(e)}$ matrix, where $N^{(e)}$ is the number of states $|m\R$ that give a nonzero contribution to the first sum above.
Likewise, The $Q^{(h)}_{\mu m}$ form a $2L\times N^{(h)}$ matrix.
Let $N=N^{(e)}+N^{(h)}$ and let us introduce a $2L\times N$ matrix $\Qv$ by joining vertically the matrix $\Qv^{(h)}$ below the matrix $
\Qv^{(e)}$, and let $\om_r$ denote the elements of the concatenated sets $\{\om^{(e)}_m\}$ and $\{\om^{(h)}_n\}$.
Then we can write
\BEQ\label{eq:qmatrix1}
G'_{\mu\nu}(\om) = \sum_r{Q_{\mu r}Q^*_{\nu r}\over \om-\om_r} 
\EEQ
If we introduce the diagonal matrix $\Lambda_{rs} = \delta_{rs}\om_r$ and
\BEQ
\gv(\om) = {1\over \om - \Lambdav} 
\EEQ
then we have the matrix expression
\BEQ\label{eq:qmatrix2}
\Gv(\om) = \Qv \gv(\om) \Qv^\dg
\EEQ
This is a very general representation of the exact cluster Green function.

\subsection{The Lehmann representation and the CPT Green function\label{subsec:Lehmann}}

Let us see how the CPT Green function can be explicitly represented in terms of the Lehmann representation (\ref{eq:qmatrix1}).
The CPT Green function (\ref{eq:CPT1}) can be written as\cite{Aichhorn:2006rt}
\BEA
\Gv(\kvt,\om) &=& {1\over (\Qv \gv(\om) \Qv^\dg)^{-1}-\Vv(\kvt)} \notag \\
&=& \Qv \gv(\om) \Qv^\dg + (\Qv \gv(\om) \Qv^\dg)\Vv(\Qv \gv(\om) \Qv^\dg) + \cdots  \notag \\
&=& \Qv \Big( \gv(\om) +  \gv(\om)(\Qv^\dg \Vv \Qv) \gv(\om) + \cdots \Big)\Qv^\dg \notag \\
&=& \Qv {1\over \om- \Lv(\kvt)} \Qv^\dg 
\EEA
where $\Lv(\kvt) = \Lambdav + \Qv^\dg\Vv(\kvt)\Qv$.
The poles of $\Gv(\kvt,\om)$ are those of $[\om- \Lv(\kvt)]^{-1}$, which we denote as $\om_r(\kvt)$.
They are simply the eigenvalue of the $N\times N$ matrix $\Lv(\kvt)$.

Let $\Uv(\kvt)$ the matrix that diagonalizes $\Lv(\kvt)$, such that
\BEQ
\Uv(\kvt)\Lv(\kvt)\Uv^\dg(\kvt) = \tilde\Lambdav(\kvt) 
\EEQ
where $\tilde\Lambdav(\kvt)$ is diagonal.
Then we write
\BEQ\label{eq:qmatrix3}
\Gv(\kvt,\om) = \Qv {1\over \om- \Lv(\kvt)} \Qv^\dg = \Qv \Uv(\kvt){1\over \om- \tilde\Lambdav(\kvt)} (\Qv \Uv(\kvt))^\dg 
\EEQ
which again is of the same form as (\ref{eq:qmatrix1}), with $\Qv$ replaced by $\tilde\Qv(\kvt) = \Qv \Uv(\kvt)$.

The representations (\ref{eq:qmatrix1}) or (\ref{eq:qmatrix3}) ensure the positivity of the cluster Green function and the CPT Green function respectively, i.e., the positive character of the corresponding spectral functions.
Indeed, the local (cluster) spectral weight is 
\BEQ
A_\mu(\om) = -2\lim_{\eta\to0}\Im G'_{\mu\mu}(\om+i\eta) 
\EEQ
and 
\BEQ
G'_{\mu\mu}(\om) = \sum_r {|Q_{\mu r}|^2 \over \om-\om_r} 
\EEQ
This expression has poles on the real axis only with positive residues, and this garantees that the corresponding spectral function $A_\mu(\om)$ is positive.
Moreover, the property $\Qv\Qv^\dg = \id$ ensures it is normalized.

The same reasoning as above applies to the CPT Green function (\ref{eq:qmatrix3}), because it has the same Lehmann structure, and the matrix $\tilde\Qv(\kvt)$ also has the property $\tilde\Qv(\kvt)\tilde\Qv(\kvt)^\dg =\id$, since the matrix $\Uv(\kvt)$ is unitary.

\section{Group theoretical concepts}
\label{section:groups}

This short appendix summarizes some key group-theoretical concepts necessary to understand the discussion of Sect.~\ref{subsec:symmetries}.
Of course this is no substitute to a text on group theory.
It merely serves as a reminder to those who have some knowledge of it, or indicates some important concepts to those who don't.

Let $\Gg$ denote the discrete symmetry group of the system and $|\Gg|$ the number of elements in the group (the order of the group).
Elements of the group will be denoted by latin letters like $g$, $h$, etc. and $gh$ will stand for group multiplication, i.e., the symmetry operation obtained by applying first $h$, then $g$.
Recall that the set $\Gg$ forms a group if the following conditions are met:
\BE
\item The set must be closed under the group multiplication, i.e., if $g_1$ and $g_2$ belong to $\Gg$, so must $g_1 g_2$.
\item There must be a neutral element $e$ (the identity transformation) such that $eg = ge$.
\item Each element $g$ must have a unique inverse $g^{-1}$ such that $gg^{-1}=g^{-1}g=e$.
\item The group operation must be associative : $(g_1g_2)g_3 = g_1(g_2g_3)$.
\EE 
The group multiplication may or may not be commutative. In the first case, the group is said to be Abelian.

The simplest non trivial group is $C_2$, the group of two elements formed by the identity transformation and a $\pi$ rotation (or, equivalently, an inversion).
Examples of cluster systems with this symmetry group are illustrated on Fig.~\ref{fig:C2}.
Another common symmetry group is $C_{2v}$, which consists of a $\pi$-rotation $c_2$ and two unequivalent reflexions ($\s_1$ and $\s_2$).
This is the symmetry group of a rectangular cluster, or of a square cluster with $d_{x^2-y^2}$ pairing, for instance.
Examples are illustrated on Fig.~\ref{fig:C2v}.

\InsertFigure{Clusters with $C_2$ symmetry. Black and grey dots represent unequivalent sites, e.g., because of the presence of a N\'eel Weiss field. The top and bottom clusters are invariant under $\pi$-rotations, and the right-most cluster is invariant under a left-right inversion.
The distinction is of course irrelevant for the middle, one-dimensional cluster.\label{fig:C2}}{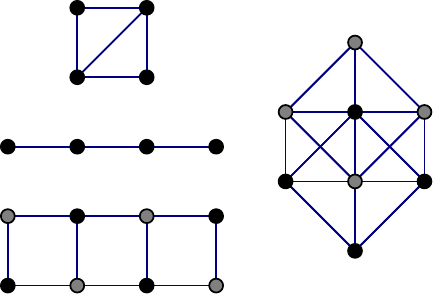}{0.6\hsize}
\InsertFigure{Clusters with $C_{2v}$ symmetry. Dashed links on the square cluster illustrate the presence of a $d_{x^2-y^2}$ pairing field, which makes horizontal and vertical links unequivalent.\label{fig:C2v}}{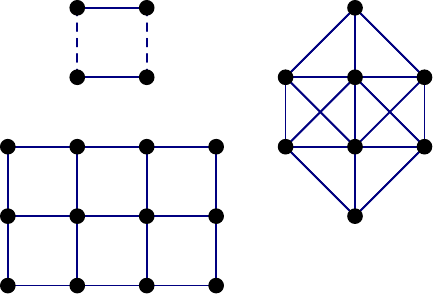}{0.6\hsize}

A \textit{representation} of the group $\Gg$ is a set of matrices that behave exactly like the group elements when group multiplication is mapped onto matrix multiplication (i.e., there is an isomorphism between the abstract group and the set of matrices).
In practice, quantum mechanics deals with group representations.
The word \textit{representation} is also applied to the vector space (or module) on which the matrix representation is based.
A representation is said to be \textit{reducible} if a change of basis can bring all group elements to the same block-diagonal form.
Thus, reducible representations are direct sums of irreducible representations.
It it the latter that are important, in great part because of Schur's lemmas, which imply that if a Hamiltonian matrix $H$ commutes with all the group elements and if the basis states are arranged into irreducible representations, then $H$ has no matrix elements between states belonging to different representations, i.e., it is block diagonal.
We often say \textit{irrep} for `irreducible representation'.

\begin{figure}
\centering 
\mbox{
\begin{tabular}{|L|RR|}
\hline
C_2 & e & c_2 \\
\hline
A & 1 & 1 \\
B & 1 & -1 \\
\hline
\end{tabular}
}
\mbox{
\begin{tabular}{|L|RRRR|}
\hline
C_{2v} & e & c_2 & \s_1 & \s_2 \\
\hline
A_1 & 1 & 1 & 1 & 1\\
A_2 & 1 & 1 &-1 &-1 \\
B_1 & 1 &-1 & 1 &-1 \\
B_2 & 1 &-1 &-1 & 1 \\
\hline
\end{tabular}
}
\vglue3mm
\begin{tabular}{|L|RRRRR|}
\hline
C_{4v} & e & c_2 & 2c_4 & 2\s_1 & 2\s_2 \\
\hline
A_1 & 1 & 1 & 1 & 1 & 1 \\
A_2 & 1 & 1 & 1 &-1 &-1 \\
B_1 & 1 & 1 &-1 & 1 &-1 \\
B_2 & 1 & 1 &-1 &-1 & 1 \\
E   & 2 &-2 & 0 & 0 & 0 \\
\hline
\end{tabular}
\caption{Character tables for the groups $C_2$, $C_{2v}$ and $C_{4v}$.\label{fig:characters}}
\end{figure}

Two group elements $g_1$ and $g_2$ are said to be \textit{conjugate} to each other if $g_1 = h^{-1}g_2h$ for some element $h$ of the group.
This property is transitive, and therefore all the elements of a group may be organized into equivalence classes called \textit{conjugacy 
classes}.
Because elements of a conjugacy class are related by a similarity transformation, they all have the same trace in a given representation.
This trace is called the \textit{character} (denoted $\chi$) of the class in the said representation.
The identity element $e$ forms a conjugacy class all by itself, and its character is the dimension of the representation.
It can be shown that the number of unequivalent irreps is the same as the number of conjugacy classes.
Characters are often displayed in tables (Fig.~\ref{fig:characters}), as a function of the conjugacy class (horizontal) and irreps (vertical).
These tables are extremely useful, for instance, to reduce tensor products of irreps.
Indeed, the trace of a matrix tensor product is the product of the traces of the factors, whereas the trace of a direct sum of matrices is the sum of the traces.

Consider, for instance, the character table of $C_{2v}$.
We learn from it that this group has 4 disinct irreps, all of dimension one.
Representation $A_1$ is the trivial representation, with states even under all symmetry transformations.
$A_2$ contains states that are odd under either reflexion.
$B_1$ and $B_2$ contain states that are odd under a $\pi$-rotation, and under one of the two reflexions.
$C_{4v}$, on the other hand, has 5 irreps, of which the last ($E$) is two-dimensional and contains states that are mixed or interchanged under $\pi/2$ rotations or reflexions.

When working in a reducible space that is the direct sum of different irreducible representations, it is possible to project onto the various irreps making up that space using the following \textit{projection operators}:
\BEQ
P^{(\a)} = \frac{d_\a}{|\Gg|}\sum_g \chi^{(\a)*}_g g
\EEQ
where $g$ stands for a symmetry operation in the reducible space considered and $\chi^{(\a)}_g$ is the character in the irrep $\a$ of the corresponding group element.

\section{Calculating averages from the Green function}\label{section:averages}

This appendix explains how to calculate the expectation value of a one-body operator from the Green function.

A general one-body term of the Hamiltonian is written as
\BEQ
\Oc = s_{\a\b}c_{\a}^\dg c_{\b}
\EEQ
where the indices $\a$ and $\b$ stand for all degrees of freedom on $\g\otimes B$: lattice site, spin and band.
In the simple case of the number of electrons, the matrix $\sv$ is diagonal:
\BEQ
s_{\rv\s,\rv'\s'} = \delta_{\rv\rv'}\delta_{\s\s'}
\EEQ
In the case of the antiferromagnetic order parameter, it is also diagonal and has the form
\BEQ
s_{\rv\s,\rv'\s'} = \delta_{\rv\rv'}\delta_{\s\s'} (-1)^\s \er^{i\Qv\cdot\rv} \qquad \Qv=(\pi,\pi)
\EEQ
We are interested in the expectation value density $\bar\Oc = s_{\a\b}\L c_\a^\dg c_\b\R/N$.

From the Lehmann representation of the Green function, we see that $\L c_\a^\dg c_\b\R$ is given by the integral of the Green function along a contour $C_<$ surrounding the negative real frequency axis counterclockwise:
\BEQ
\L c^\dg_\b c_\a \R = \int_{C_<}{\dr z\over 2\pi i} G_{\a\b}(z)
\EEQ
Therefore the expectation value we are looking for is
\BEQ
\bar\Oc = \frac1N s_{\b\a}\L c^\dg_\b c_\a \R = \frac1N \int_{C_<}{\dr z\over 2\pi i} \tr\l[\sv\Gv(z)\r] 
\EEQ
Here the trace includes a sum over lattice sites, spin and band.
In a mixed representation, with cluster sites indices and reduced wavevector instead of the lattice site index, this becomes
\BEQ\label{eq:expect1}
\bar\Oc  = \frac1N\sum_\kvt \int_{C_<}{\dr z\over 2\pi i} \tr\l[\sv(\kvt)\Gv(\kvt,z)\r] 
\EEQ
where we assumed that the matrix $\sv$ is diagonal in $\kvt$.

Next, let us consider the asymptotic behavior of the Green function as $z\to\infty$: $\Gv(z)\to \id/z$.
This allows us to modify Eq.~(\ref{eq:expect1}) as follows:
\BEQ
\bar\Oc  = \frac1N\sum_\kvt \int_{C_<}{\dr z\over 2\pi i} \l\{ \tr\l[\sv(\kvt)\Gv(\kvt,z)\r] -\frac{\tr\sv(\kvt)}{z-p}\r\}
\EEQ
where $p>0$ (in practice, we use $p\sim 1$).
The term we added does not contribute, since its unique pole lies outside of the contour.
However, this term modifies the asymptotic behavior of the integrand, which nows decays as $1/z^2$.
This allows us to replace the contour $C_<$ by an integral along the imaginary axis, plus an infinite semi-circle that does not contribute, since the integrand falls faster than $1/z$.

Next, consider the part of the contour $C_<$ that lies above the real axis, and let us follow this contour clockwise and call it $C$. Let $C'$ be the mirror image of $C$ below the real axis, followed counterclockwise.
To each $z$ and $\dr z$ of $C$ correspond the mirror images $z^*$ and $\dr z^*$ on $C'$, so that
\BEQ
I[C'] = \int_{C'} \dr z f(z) = \int_C \dr z^* f(z^*)
\EEQ
If, in addition, the integrand is such that $f(z^*) = f^*(z)$, then
\BEQ
I[C'] = \int_C \dr z^* f^*(z) =\l(\int_C \dr z f(z)\r)^* = I^*[C] 
\EEQ
The integral of $f(z)$ along the counterclockwise contour $C_<$ would then be
\BEQ
I[C_<] = I[C'] - I[C] = I^*[C] - I[C] = -2i\Im I[C]
\EEQ
One of the properties of the Green function is its hermiticity: $G_{\a\b}(z^*) = G_{\b\a}^*(z)$.
In the mixed Fourier representation, this is rather expressed as $\Gv(\kvt,z^*) = \Gv^\dg(-\kvt,z)$.
We also assume that $\sv$ is Hermitian: $\sv(\kvt) = \sv^\dg(-\kvt)$ so that the expectation value is real.
This means that the integrand of the expectation value respects the condition $f(z^*) = f^*(z)$.

Finally, the expectation value has the expression 
\BEQ\label{eq:expect2}
\bar\Oc  = \frac1N\sum_\kvt \int_0^\infty {\dr\om\over\pi} \Re \l\{ \tr\l[\sv(\kvt)\Gv(\kvt,i\om)\r] -\frac{\tr\sv(\kvt)}{i\om-p}\r\}
\EEQ

\section{Evaluation of the Potthoff functional}\label{section:frequencyIntegral}

In this appendix, we show how to evaluate the frequency integral (\ref{eq:freqInt}).
Let us take that contour to be the whole imaginary axis between $-iR$ and $iR$, with a half-circle of radius $R$ closing the contour on the left part of the complex plane.
Let us first see what the behavior of the integrand is as $|\om|\to\infty$.
Since $\Gv'(\om) \sim 1/\om$ as $\om\to\infty$, one may write, in that limit,
\ALIGN{
\ln\det\Big[&1-\Vv(\kvt)\Gv'(\om)\Big] = \tr\ln\Big[1-\Vv(\kvt)\Gv'(\om)\Big] \notag \\
&\sim \tr\ln\Big[1-{\Vv(\kvt)\over\om}\Big] \sim -{1\over\om}\tr\Vv(\kvt) 
}
The integration over wavevectors yields
\BEQ
\frac{L}{N}\sum_\kvt \tr\Vv(\kvt) = \frac{L}{N}\sum_\a t_{\a\a} - \sum_\mu t'_{\mu\mu}\quad (\om\to\infty) 
\EEQ
The only contribution from these terms is the chemical potential, by definition.
Thus the $1/\om$ term in the integral $I$ is
\BEQ
P(\om) = 2L(\mu-\mu'){1\over\om} 
\EEQ
This term by itself does not contribute to the frequency integral.
Indeed, let us arrange for the contour $C$, which normally should cross the origin, to avoid it along an infinitesimal semi-circle $C_1$ of radius $\eta$ centered at the origin and lying on the left-hand half-plane.
This slight modification should not change the value of $\Om$, if we refer to the exact method of the previous subsection, as the zero frequency does not contribute.
Then the above term does not contribute, since the pole at $\om=0$ lies outside the contour.
It can therefore be subtracted and the frequency integral has the following expression:
\[
I =  \int_C{\dr\om\over 2\pi i} \l\{\frac{L}{N}\sum_\kvt\ln\det(1-\Vv(\kvt)\Gv'(\om)) - P(\om)\r\}
\]
in which the integrand now falls like $1/\om^2$ at large frequencies, and the half-circle of radius $R\to\infty$ does not contribute to the integral.
Let us use the properties
\BEQ
\Gv(-ix) = \Gv(ix)^* \Quad \Vv(-ix,\kvt) = \Vv(ix,-\kvt)^*
\EEQ
to express the integral over the whole imaginary axis as an integral over the `positive' imaginary axis only. With $\om=ix$, we write
\SPLIT{
I = &\int_\eta^R {\dr x\over 2\pi} \frac{L}{N}\sum_\kvt\Big[
\ln\det(1-\Vv(\kvt)\Gv'(ix)) \\  & + \ln\det(1-\Vv(-\kvt)^*\Gv'(ix)^*)\Big]
- \int_{C_1}{\dr\om\over 2\pi i}P(\om) 
}
Note that $P(\om)$ does not contribute to the integral along the imaginary axis, since it is odd in $x$; in other words, the principal value is taken and corresponds to the contribution of the small half-circle $C_1$.
We also note that the main part of the integrand does not have a contribution along the contour $C_1$, since the logarithmic singularity near $\om=0$ is integrable, i.e., leads to a vanishing contribution on $C_1$ as $\eta\to0$.
Since integrating over $\kvt$ and over $-\kvt$ are equivalent, one further simplifies to
\BEQ
I = \int_0^\infty {\dr x\over\pi} \frac{L}{N}\sum_\kvt  \ln\Big|\det(1-\Vv(\kvt)\Gv'(ix))\Big| - L(\mu-\mu')
\EEQ

The frequency integral is done by dividing the positive imaginary frequency axis in three segments: $[0,\Lambda_1]$, $[\Lambda_1,\Lambda_2]$ and $[\Lambda_2,\infty)$.
The constant $\Lambda_1$ is a low-energy scale in the problem, like the lowest eigenvalue $\om'_r$ (up to some minimum), whereas $\Lambda_2$ is a high-energy scale, like the largest eigenvalue $\om'_r$ (up to some maximum).
On each segment a 20-point Gauss-Legendre integration is used.
The last segment is in fact treated like an integral over $u=1/\om$ from 0 to $\Lambda_2^{-1}$.
The frequency integral $\int \dr\om\,f(\om)$ thus takes the form of a weighted sum $\sum_n p_n f(\om_n)$, where $f(\om)$ is the wavevector sum conducted at frequency $\om$.
The required accuracy of the wavevector sum, which sets the number of points of the wavevector mesh, is conditioned by the weight $p_n$ (i.e. it does not have to be so large when $p_n$ is small).
In addition, the integrand of (\ref{eq:sef4}) may have sharp structures -- thus requiring a fine wavevector mesh -- at frequencies close to the real axis (the poles of $\Gv$ are all real) but is increasingly smooth as one moves away from the real axis.
Thus the wavevector mesh may be redefined from time to time as one progresses along the imaginary frequency axis, making the mesh coarser and coarser without cost in accuracy.

Tests have been conducted in order to compare the speed and accuracies of the two methods: the analytic frequency integration (AI) described in Sect.~\ref{subsec:AI}, and the numerical frequency integration (NI) described in this section.
In each case, a fixed mesh and an adaptive mesh have been used.
The results are displayed in Figs.~\ref{fig:precision1} and \ref{fig:precision2}.

\InsertFigure{Comparisons of the different integration methods in execution time and accuracy for a 4-site cluster. See text for details.\label{fig:precision1}}{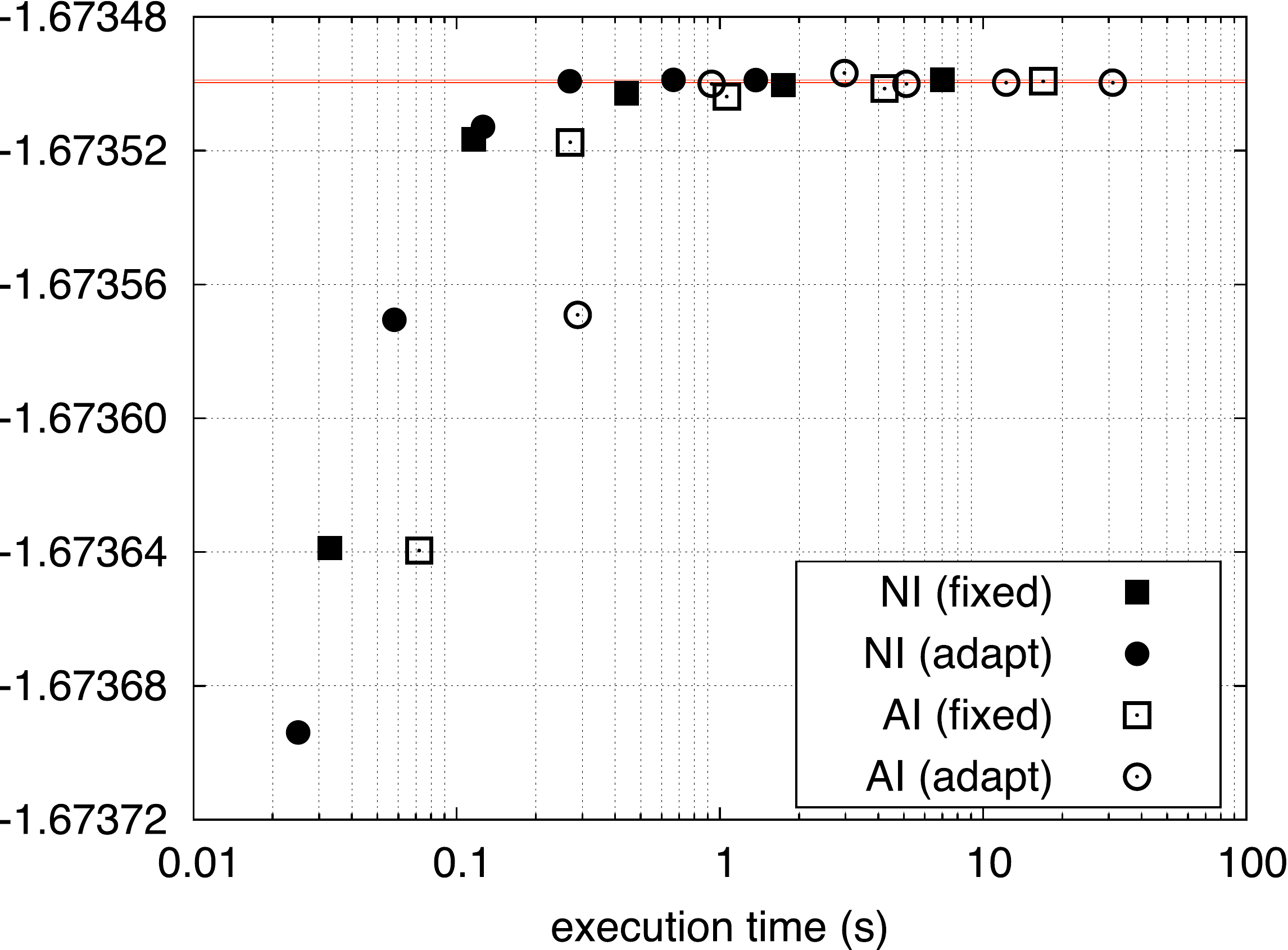}{0.9\hsize}
Fig.~\ref{fig:precision1} shows the relation between execution time (in seconds) on a 2.16 MHz Intel Core 2 Duo processor and the value of the functional obtained with four different methods: (i) Numerical integration (NI) with a fixed wavevector grid, (ii) NI with an adaptive wavevector grid, (iii) Analytic integration (AI) with a fixed wavevector grid and (iv) AI with an adaptive wavevector grid. For each method, different points correspond to different required accuracies of the wavevector integral.
The system used in that case was the two-dimensional Hubbard model with nearest-neighbor hopping $t=1$, Coulomb repulsion $U=8$ and chemical potential $\mu=1$, on a 4-site cluster.
The time to perform the exact diagonalization is negligible here.
The value of the chemical potential used makes the system metallic, i.e., there are poles of the VCA Green function at zero frequencies, which is more demanding on the integral because of sharp features of the integrand.
Deep within ordered phases there are generally gaps in the physical spectrum that make the integrals converge faster, but the location of phase boundaries, where these gaps disappear, is generally of great interest.
One sees from Fig.~\ref{fig:precision1} that the NI method with adaptive mesh converges fastest, about 3 times faster than the AI method.
The two horizontal red lines represent the two converged values (for NI and AI).
They differ by less than $10^{-6}$, a difference due to the finite mesh used in the frequency integral.
This accuracy is more than adequate for applications.

\InsertFigure{Same as Fig.~\ref{fig:precision1} for a 12-site cluster.\label{fig:precision2}}{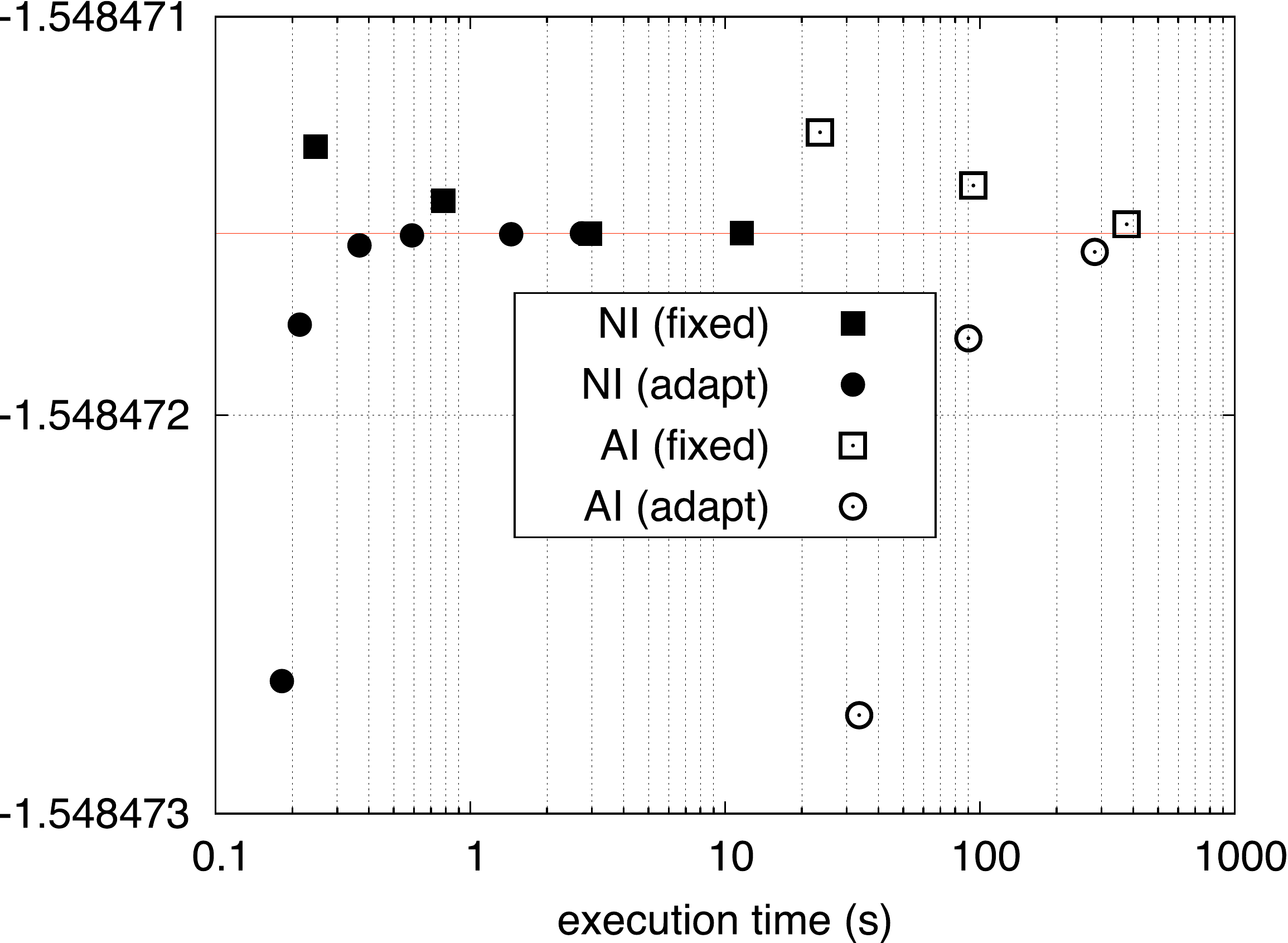}{0.9\hsize}

Fig.~\ref{fig:precision2} shows the same, this time for a 12-site cluster.
The exact diagonalization is done beforehand, and is the same for all integration methods shown.
In that case the number $R$ of poles was 312, instead of $R=32$ for the 4-site cluster used in Fig.~\ref{fig:precision1}.
Here the gain in using NI is greater even, as the method is over 500 times faster than AI with a adaptive mesh.
Indeed, it is prohibitively expensive to use the AI method for anything but very small clusters.
Method I (AI) is of order $N_kR^3$, whereas Method II (NI) involves linear-algebraic operations on the Green function, of order $N_k N_\om L^3$ (say, for a fixed wavevector grid and $N_\om$ frequencies).
The ratio of execution times between the two method should roughly be
\BEQ
\frac{\text{time(AI)}}{\text{time(NI)}} \sim \l(\frac{R}{L}\r)^3 \frac{1}{N_\om}
\EEQ
In the case of Fig.~\ref{fig:precision1}, $R=32$, $L=4$ and $N_\om=60$, which gives a ratio of $\sim 8$.
In the case of Fig.~\ref{fig:precision2}, $R=312$, $L=12$ and $N_\om=60$, which gives a ratio of $\sim 290$.
In both cases this overestimates by a factor 2 to 3 the measured advantage of Method II, which is nevertheless enormous.

\section{Fermionic Baths and hybridization functions}\label{section:bath}

In this short appendix we consider the Hamiltonian \BEQ
H = \sum_{\mu\nu} t_{\mu\nu}c_\mu^\dg c_\nu + \sum_\a \eps_\a a^\dg_\a a_\a 
+ \sum_{\mu,\a} \l(\t_{\mu\a} c^\dg_\mu a_\a  + \t^*_{\mu\a} a^\dg_\a c_\mu\r)
\EEQ
and show that the Green function obtained by tracing over the bath degrees of freedom has the form \BEQ
(\Gv^{-1})_{\mu\nu} = \om -t_{\mu\nu} - \sum_\a \frac{\t_{\mu\a}\t^*_{\nu\a}}{\om-\eps_\a}~.
\EEQ

First of all, the full Green function associated with the above one-body Hamiltonian is 
\BEQ
\Gv_{\rm full}(\om) = \frac1{\om-\Tv}
\EEQ
where the full hopping matrix has the block form \BEQ
\Tv = \MATRIX{- \tv & \thetav \\ \thetav^\dg & -\epsv}
\EEQ
where $\tv$ is the hopping matrix within cluster degrees of freedom only, $\thetav$ the hopping matrix between bath and cluster orbitals, and $
\epsv$ the diagonal matrix of bath energies $\eps_\a$.
The Green function obtained by tracing out the bath degrees of freedom is simply the restriction of $\Gv_{\rm full}$ (and not of its inverse) to the cluster degrees of freedom only.
The mathematical problem at hand is simply to invert a $2\times2$ block matrix 
\BEQ
\MATRIX{A_{11} & A_{12} \\ A_{21} & A_{22}} = \MATRIX{B_{11} & B_{12} \\ B_{21} & B_{22}}^{-1}~,
\EEQ
where $A_{11}=\om-\tv$, $A_{12} = A_{21}^\dg = \thetav$, $A_{22}=\om-\epsv$, and $B_{11}$ is the Green function we are looking for.
By working out the inverse matrix condition, we find in particular that \ALIGN{
A_{11}B_{11} + A_{12}B_{21} = \id \\
B_{21} = -A_{22}^{-1}A_{21}B_{11}
}
and therefore 
\BEQ
\l( A_{11} - A_{12}A_{22}^{-1}A_{21}\r) B_{11} = \id~.
\EEQ
The Green function is thus
\ALIGN{
\Gv^{-1} &= \om - \tv - \thetav\frac1{\om-\epsv}\thetav^\dg \notag\\
&= \om - \tv - \Gammav(\om)
}
where we defined the so-called \textit{hybridization function}
\BEQ
\Gamma_{\mu\nu}(\om) = \sum_\a \frac{\t_{\mu\a}\t^*_{\nu\a}}{\om-\eps_\a}
\EEQ
Note that nowhere but in the last expression have we supposed that the matrix $\epsv$ is diagonal.
That condition simply serves to minimize computation time.

%
\end{document}